\documentclass[12pt,a4paper]{iopart}

\usepackage{iopams}
\usepackage{setstack}
\usepackage{graphicx}
\usepackage{pict2e}
\usepackage{color}
\usepackage{hyperref}

\renewcommand{\P}{\mathbb{P}}
\newcommand{\C}[2]{{#1 \choose #2}}
\newcommand{\Li}{\mathrm{Li}}
\newcommand{\Ai}{\mathrm{Ai}}
\newcommand{\Bi}{\mathrm{Bi}}
\newcommand{\erf}{\mathrm{erf}}
\renewcommand{\Re}{\mathrm{Re}}

\newcommand{\openone}{{\bf1}}
\newcommand{\rhobar}{\overline{\rho}}
\newcommand{\tm}{t_{\text{m}}}
\newcommand{\te}{t_{\text{e}}}

\newcommand{\rhom}{\rho_{\text{m}}}
\newcommand{\rhoe}{\rho_{\text{e}}}

\newcommand{\He}{\mathcal{H}_{\text{e}}}

\newcommand{\bb}{\text{bb}}

\newcommand{\keywords}[1]{\noindent\textbf{Keywords:} #1}

\renewcommand{\text}[1]{\mathrm{#1}}
\newtheorem{conjecture}{Conjecture}

\newcounter{X}
\newcounter{Y}
\newcommand{\up}{\put(\theX,\theY){\line(1,1){10}\put(2,6){$E$}}\addtocounter{X}{10}\addtocounter{Y}{10}}
\newcommand{\down}{\put(\theX,\theY){\line(1,-1){10}\put(5,-4){$D$}}\addtocounter{X}{10}\addtocounter{Y}{-10}}
\newcommand{\hz}{\put(\theX,\theY){\line(1,0){10}\put(4,1){$\openone$}}\addtocounter{X}{10}}

\begin{document}

\title{Brownian bridges for late time asymptotics of KPZ fluctuations in finite volume}
\author{Kirone Mallick$^1$, Sylvain Prolhac$^2$}
\address{$^1$Institut de Physique Th\'eorique, CEA, CNRS-URA 2306, Gif-sur-Yvette, France}
\address{$^2$Laboratoire de Physique Th\'eorique, IRSAMC, UPS, Universit\'e de Toulouse, France}


\begin{abstract}
Height fluctuations are studied in the one-dimensional totally asymmetric simple exclusion process with periodic boundaries, with a focus on how late time relaxation towards the non-equilibrium steady state depends on the initial condition. Using a reformulation of the matrix product representation for the dominant eigenstate, the statistics of the height at large scales is expressed, for arbitrary initial conditions, in terms of extremal values of independent standard Brownian bridges. Comparison with earlier exact Bethe ansatz asymptotics leads to explicit conjectures for some conditional probabilities of non-intersecting Brownian bridges with exponentially distributed distances between the endpoints.\\

\end{abstract}


\begin{section}{Introduction and summary of main results}
Non-equilibrium settings involving many degrees of freedom in interaction have been increasingly investigated in the past decades. A particularly interesting and much studied \cite{KK2010.1,TSSS2011.1,QS2015.1,HHT2015.1,S2016.2} emergent dynamics at large scales, known as KPZ universality \cite{KPZ1986.1}, describes systems such as growing interfaces or driven particles with short range dynamics and enough non-linearity. A central object in KPZ universality is the height function $h$, a random field depending on space and time. Once the initial condition $h_{0}$ and the geometry (infinite system, finite system with periodic boundaries, open system in contact with the environment at the boundaries, \ldots) are specified, the statistics of $h$ is uniquely determined for all systems in KPZ universality.

In one dimension, several exactly solvable models belong to KPZ universality. The study of these models has led in the past twenty years to a precise characterization of the statistics of the height $h(x,t)$ in a variety of settings. On the infinite line, connections to statistics of extremal eigenvalues in random matrix theory (Tracy-Widom distributions and Airy processes) have been discovered \cite{J2005.1,S2006.1,S2007.1,F2010.1,C2011.1}. More recently \cite{P2016.1,BL2017.1,L2016.1,BL2017.2}, some progress was made for periodic boundaries with some specific initial conditions. There, finite volume leads to quantization of allowed momenta $k$ in Fourier space, and the relaxation modes are described by elementary excitations of particle-hole type \cite{P2014.1} with dispersion $k^{3/2}$ \cite{F2002.1}, which can be analyzed by Bethe ansatz.

In this work, we consider the long time limit of one-dimensional KPZ fluctuations $h(x,t)$ with periodic boundaries $x\equiv x+1$ and arbitrary initial condition. Starting with a recent matrix product characterization \cite{LM2011.1,L2013.1} of the dominant eigenvector of the totally asymmetric simple exclusion process (TASEP) \cite{D1998.1,S2001.1,GM2006.1}, an exactly solvable model in KPZ universality which features hard-core particles hopping in the same direction, we express the long time statistics of $h(x,t)$ with initial condition $h_{0}$ in terms of extremal values of independent Brownian bridges (see section \ref{section theta bb}). Equivalently, our results are formulated, in section \ref{section non intersecting}, as conditional probabilities of \textit{non-intersecting} Brownian bridges with exponentially distributed distances between the endpoints. Comparison with earlier results \cite{P2016.1} for specific initial conditions leads to precise conjectures for these Brownian functional (section \ref{section conjectures}). We also perform perturbative expansions for more general initial conditions. In particular, we find, in section \ref{section cumulants}, that when the amplitude of the initial height $h_{0}$ is small, the average height is given in the long time limit by
\begin{equation*}
\fl\hspace{5mm}
\langle h(x,t)\rangle-t\simeq\frac{\sqrt{\pi}}{2}-\frac{\sqrt{2\pi}}{4}+\int_{0}^{1}\!\rmd x\,h_{0}(x)+\sqrt{2\pi}\,\sum_{k\in\mathbb{Z}}a_{k}a_{-k}(-1)^{k}k\pi J_{1}(k\pi)+\ldots
\end{equation*}
where the $a_{k}$'s are the Fourier coefficients of the initial height $h_{0}$.

The paper is organized as follows. In the rest of this section, known facts about the relation between TASEP and KPZ universality are recalled and our main results are precisely stated. In section \ref{section matrix product -> bb}, we derive these results by rewriting the matrix product expressions of \cite{LM2011.1,L2013.1} in a simpler form, which allows us to take the limit of large system sizes in a straightforward manner. In section \ref{section expectation values bb}, we calculate various Brownian expectation values related to the first cumulants of $h(x,t)$. Some technical results are gathered in the appendices.

\begin{subsection}{Periodic TASEP}
At the microscopic level, we consider the TASEP dynamics on a one-dimensional lattice (Fig. \ref{fig TASEP}). Each site is either empty or occupied by a single particle, and each particle may hop independently of the others from its current site $i$ to the site $i+1$ if the latter is empty. The hopping rate is fixed equal to $1$ with respect to the (microscopic) time scale $\tm$.

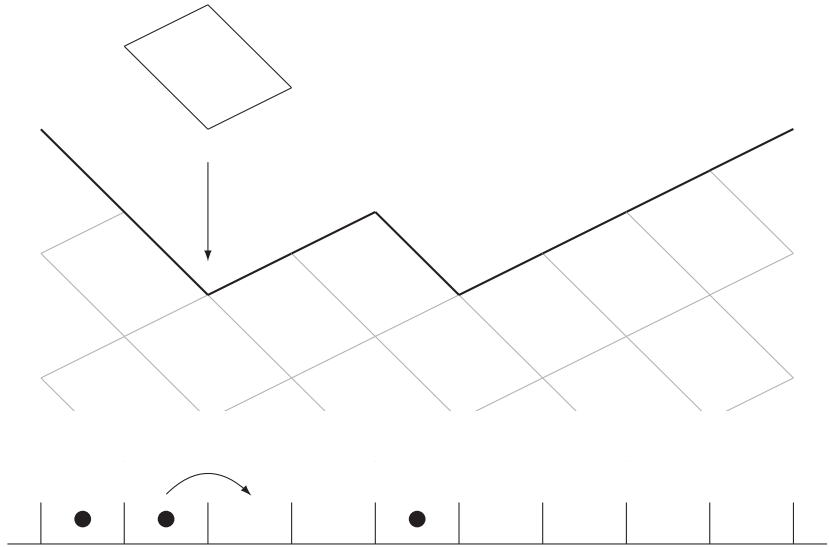
\begin{figure}
  \begin{center}
  \begin{tabular}{c}
    \newcommand{\Up}[1]{\put #1{\line(1,0.5){10}}}
    \newcommand{\Down}[1]{\put #1{\line(1,-1){10}}}
    \newcommand{\UpThick}[1]{\put #1{\thicklines\line(1,0.5){10}}}
    \newcommand{\DownThick}[1]{\put #1{\thicklines\line(1,-1){10}}}
    \newcommand{\UpDotted}[1]{\put #1{\color[rgb]{0.7,0.7,0.7}\line(1,0.5){10}\color{black}}}
    \newcommand{\DownDotted}[1]{\put #1{\color[rgb]{0.7,0.7,0.7}\line(1,-1){10}\color{black}}}
    \setlength{\unitlength}{1.1mm}
    \begin{picture}(104,70)(6,0)
    \put(15,3){\circle*{2}}
    \put(25,3){\circle*{2}}
    \put(55,3){\circle*{2}}
    \qbezier(25,6)(30,11)(35,6)\put(35,6){\vector(1,-1){0.2}}
    \put(6,0){\line(1,0){98}}
    \multiput(10,0)(10,0){10}{\line(0,1){5}}
    \DownDotted{(10,20)}\UpDotted{(10,20)}\DownDotted{(10,35)}\UpDotted{(10,35)}\DownThick{(10,50)}
    \UpDotted{(20,10)}\DownDotted{(20,25)}\UpDotted{(20,25)}\DownThick{(20,40)}
    \UpDotted{(30,15)}\DownDotted{(30,30)}\UpThick{(30,30)}
    \DownDotted{(40,20)}\UpDotted{(40,20)}\DownDotted{(40,35)}\UpThick{(40,35)}
    \UpDotted{(50,10)}\DownDotted{(50,25)}\UpDotted{(50,25)}\DownThick{(50,40)}
    \UpDotted{(60,15)}\DownDotted{(60,30)}\UpThick{(60,30)}
    \DownDotted{(70,20)}\UpDotted{(70,20)}\DownDotted{(70,35)}\UpThick{(70,35)}
    \UpDotted{(80,10)}\DownDotted{(80,25)}\UpDotted{(80,25)}\DownDotted{(80,40)}\UpThick{(80,40)}
    \UpDotted{(90,15)}\DownDotted{(90,30)}\UpDotted{(90,30)}\DownDotted{(90,45)}\UpThick{(90,45)}
    \put(5,10){\color{white}\polygon*(0,0)(100,0)(100,6)(0,6)}
    \put(30,46){\vector(0,-1){12}}
    \Down{(20,60)}
    \Up{(30,50)}
    \Up{(20,60)}
    \Down{(30,65)}
    \end{picture}
  \end{tabular}
  \end{center}
  \caption{Representation of the dynamics of TASEP in terms of hopping particles (bottom) and corresponding height function $H_{i}$ (top) for a system with density $\rhobar=1/3$.}
  \label{fig TASEP}
\end{figure}

Since we focus on KPZ fluctuations in a finite volume, we restrict in the following to a periodic lattice with $L\gg1$ sites for the microscopic model. The sites are numbered by $i\in\mathbb{Z}$, $i\equiv i+L$, with an origin $i=1$ chosen arbitrarily. The total number of particles $N$ and the average density $\rhobar=N/L$ are conserved by the dynamics. At time $\tm$, the configuration of TASEP can be specified either by the positions of the particles $X_{j}(\tm)$, $j=1,\ldots,N$, distinct modulo $L$, or by the occupation numbers $n_{i}(\tm)$, $i=1,\ldots,L$ with $n_{i}=1$ if site $i$ is occupied and $n_{i}=0$ otherwise. A more precise characterization of the state of the model is through the height function $H_{i}(\tm)$, initially equal to $H_{i}(0)=\sum_{k=1}^{i}(\rhobar-n_{k}(0))$, $i=1,\ldots,L$, and which increases by $1$ each time a particle moves from site $i$ to site $i+1$. Then, $H_{i}(\tm)=H_{0}(\tm)+\sum_{k=1}^{i}(\rhobar-n_{k}(\tm))$ at any time. The time-integrated current $Q_{i}(\tm)$ between sites $i$ and $i+1$, defined as the total number of particles that have hopped from $i$ to $i+1$ between time $0$ and $\tm$, is equal to $Q_{i}(\tm)=H_{i}(\tm)-H_{i}(0)$.

In the continuum $L\gg1$, density profiles $\rho(x)$ can be defined as local averages of the occupation numbers $n_{i}$ over a number $L\,\rmd x$ of consecutive sites around site $i=\lfloor L\,x\rfloor$, with $1\ll L\,\rmd x\ll L$. The density profile corresponding to the initial state of the microscopic model will be denoted by $\rho_{0}$ in the following. In this work, we are interested in the large scale fluctuations of the height function $H_{i}(\tm)$ in the KPZ time scale $\tm\sim L^{3/2}$, and more specifically in the precise dependency on $\rho_{0}$ of the fluctuations at late time $\tm/L^{3/2}\gg1$ when the system approaches its non-equilibrium steady state.
\end{subsection}

\begin{subsection}{Large scale dynamics in finite volume}
 We first summarize a few known results about deterministic hydrodynamics on the Euler time scale $\tm\sim L$. Then we consider finite volume fluctuations on the KPZ time scale $\tm\sim L^{3/2}$. In the latter case, the evolution starting from a typical fluctuation and the evolution starting from a finite density profile have to be treated separately.

\begin{subsubsection}{Hydrodynamics on the Euler time scale $\tm\sim L$.}
\label{section Burgers}
\hfill\break
The hydrodynamic time $\te$ is defined by $\tm=\te L$. For fixed $\te$ and average density $\rhobar$, the limit $\rhoe(x,\te)=\lim_{L\to\infty}\rhom(x,L\te)$ is well defined and equal to the viscosity solution of the inviscid Burgers' equation \cite{S1991.1}
\begin{equation}
\label{Burgers eq}
\frac{\rmd}{\rmd\te}\,\rhoe(x,\te)+\frac{\rmd}{\rmd x}\,[\rhoe(x,\te)(1-\rhoe(x,\te))]=0\; .
\end{equation}
 Burgers' equation is deterministic and conserves the total density $\int_{0}^{1}\rmd x\,\rhoe(x,\te)=\rhobar$. The initial condition $\rhoe(x,0)=\rho_{0}(x)$ and periodic boundary conditions $x\equiv x+1$ are imposed.

The height profile defined from the TASEP height function as
\begin{equation}
\He(x,\te)=\lim_{L\to\infty}L^{-1}H_{\lfloor Lx\rfloor}(L\te)
\end{equation}
is given in terms of the solution $\rhoe$ of Burgers' equation by
\begin{equation}
\label{He[rhoe]}
\He(x,\te)=\mathcal{H}_{0}(x)+\int_{0}^{\te}\rmd\tau\rhoe(x,\tau)(1-\rhoe(x,\tau))\;,
\end{equation}
with initial condition
\begin{equation}
\label{H0[rho0]}
\mathcal{H}_{0}(x)=\int_{0}^{x}\rmd y\,(\rhobar-\rho_{0}(y))\;.
\end{equation}
This height profile verifies
\begin{equation}
\hspace{-10mm}
\frac{\rmd}{\rmd x}\,\He(x,\te)=\rhobar-\rhoe(x,\te)
\quad\text{and}\quad 
\frac{\rmd}{\rmd\te}\,\He(x,\te)=\rhoe(x,\te)(1-\rhoe(x,\te))\;. 
\end{equation}

An important feature of hyperbolic conservation laws such as (\ref{Burgers eq}) is the presence of shocks (i.e. discontinuities in $x$ for $\rhoe(x,\te)$), whose number evolves in time by a complicated process of spontaneous generation and merging.
For smooth enough initial condition $\rho_{0}$, the number of shocks does not change any more after some time, since all the remaining shocks move with the same asymptotic velocity $1-2\rhobar$ in the long time limit. Surprisingly, it is not necessary to solve the whole time evolution to determine how many shocks survive when $\te\to\infty$: Theorem 11.4.1 of \cite{D2010.3} asserts that this asymptotic number of shocks is equal to the number of times the global minimum of $\mathcal{H}_{0}$ is reached.
Denoting by $\kappa$ any position where $\mathcal{H}_{0}(\kappa)$ is equal to its global minimum $\min[\mathcal{H}_{0}]$, the following long time asymptotics (in the moving reference frame with velocity $1-2\rhobar$) is satisfied:
\begin{equation}
\label{He(te) asymptotics}
\He(x+(1-2\rhobar)\te,\te)
\simeq\rhobar(1-\rhobar)\te
+\min[\mathcal{H}_{0}]
+\frac{(x-\kappa)^{2}}{4\te}\;, 
\end{equation}
for any $x$ such that there is no shock between $x$ and $\kappa$.
\end{subsubsection}

\begin{subsubsection}{KPZ time scale $t_{\text{m}}\sim L^{3/2}$: Evolution from a typical fluctuation.}
\hfill\break
We define the rescaled time $t$ by
\begin{equation}
\label{tKPZ}
\tm=\frac{tL^{3/2}}{\sqrt{\rhobar(1-\rhobar)}}
\end{equation}
and consider an initial density profile of the form
\begin{equation}
\label{rho0 fluct}
\rho_{0}(x)=\rhobar+\sqrt{\rhobar(1-\rhobar)}\;\frac{\sigma_{0}(x)}{\sqrt{L}}
\end{equation}
with $\sigma_{0}$ periodic of period $1$ such that $\int_{0}^{1}\rmd x\,\sigma_{0}(x)=0$. This corresponds to a typical fluctuation in the stationary state, such that the density profile remains almost surely equal to the constant $\rhobar$ at leading order in $L$ for any time $t$. The corresponding height function behaves at large $L$ for fixed $x$, $t$ and $\rhobar$ as
\begin{equation}
\label{H[h] fluct}
H_{(1-2\rhobar)\tm+xL}(\tm)\simeq\rhobar(1-\rhobar)\tm+\sqrt{\rhobar(1-\rhobar)L}\,h(x,t)\;.
\end{equation}
The site $i$ at which the height $H_{i}(\tm)$ is considered in (\ref{H[h] fluct}) moves at the velocity $1-2\rhobar$ of density fluctuations. The first term in the right hand side is the contribution of the instantaneous current $\rhobar(1-\rhobar)$. The height fluctuation $h(x,t)$ is a random variable; its initial value $h(x,0)=h_{0}(x)$ is given in terms of the initial density profile of TASEP by 
\begin{equation}
\label{h0[sigma0]}
h_{0}(x)=-\int_{0}^{x}\rmd y\,\sigma_{0}(y)\;.
\end{equation}
The function $h_{0}$ is continuous and verifies $h_{0}(0)=h_{0}(1)=0$.

The height fluctuation $h(x,t)$ is a random field that has the same law as a specific solution of the Kardar-Parisi-Zhang (KPZ) equation in the limit of strong non-linearity. More precisely, one has \cite{BG1997.1,DM1997.1}
\begin{equation}
\label{h[hKPZ] fluct}
h(x,t)=\lim_{\lambda\to\infty}\Big(\mathfrak{h}_{\lambda,1}(x,t/\lambda)-\frac{\lambda^{2}t}{3}\Big)
=\lim_{\ell\to\infty}\frac{\mathfrak{h}_{1,\ell}(x\ell,t\ell^{3/2})-t\ell^{3/2}/3}{\sqrt{\ell}}\;,
\end{equation}
where $\mathfrak{h}_{\lambda,\ell}(x,t)$ is the properly renormalized \cite{H2013.1} solution of the KPZ equation
\begin{equation}
\partial_{t}\mathfrak{h}_{\lambda,\ell}(x,t)=\frac{1}{2}\partial_{x}^{2}\mathfrak{h}_{\lambda,\ell}(x,t)-\lambda(\partial_{x}\mathfrak{h}_{\lambda,\ell}(x,t))^{2}+\eta(x,t)
\end{equation}
with initial condition $\mathfrak{h}_{\lambda,\ell}(x,0)=\sqrt{\ell}\,h_{0}(x/\ell)$. The Gaussian white noise $\eta$ in the KPZ equation has covariance $\langle\eta(x,t)\eta(x',t')\rangle=\delta(x-x')\delta(t-t')$, and periodic boundary conditions $\mathfrak{h}_{\lambda,\ell}(x,t)=\mathfrak{h}_{\lambda,\ell}(x+\ell,t)$ are imposed. The second equality in (\ref{h[hKPZ] fluct}) results from the scaling properties of the KPZ equation, $\mathfrak{h}_{\lambda,\ell}(x,t)=\sqrt{\alpha}\,\mathfrak{h}_{\lambda\sqrt{\alpha},\ell/\alpha}(x/\alpha,t/\alpha^{2})$ in law for arbitrary $\alpha>0$.
\end{subsubsection}

\begin{subsubsection}{KPZ time scale $t_{\text{m}}\sim L^{3/2}$: Evolution from a finite density profile.}
\hfill\break
We consider now the case of an initial state corresponding to a non-constant density profile $\rho_{0}(x)\neq\rhobar$ on the KPZ time scale (\ref{tKPZ}). Exact results \cite{P2016.1,BL2017.1} for domain wall initial condition $\rho_{0}(x)=1_{\{0\leq x\leq\rhobar\}}$ suggest the existence of an analogue to (\ref{H[h] fluct}). As far as we know, a theorem for general $\rho_{0}$ is however still missing. From (\ref{He(te) asymptotics}), an additional deterministic shift $L\min[\mathcal{H}_{0}]$ is contributed to the height by the whole Euler time scale, $\te\in\mathbb{R}^{+}$. Even though other modifications may be needed, the exact result for domain wall initial condition and simulations in a few other cases suggest that the height fluctuations are given at large $L$ with fixed $x$, $t$ and $\rhobar$ by the rather minimal modification
\begin{equation}
\label{H[h] finite}
H_{(1-2\rhobar)\tm+xL}(\tm)\simeq\rhobar(1-\rhobar)\tm+\min_{i}H_{i}(0)+\sqrt{\rhobar(1-\rhobar)L}\,\tilde{h}(x,t)\;.
\end{equation}
The tilde is used for height fluctuations in (\ref{H[h] finite}) in order to distinguish from $h(x,t)$ in (\ref{H[h] fluct}). Given the somewhat light evidence given here for (\ref{H[h] finite}), a proof would be very much welcome. In particular, it may be possible that additional shifts to the average height will be needed for general initial condition. Comparison between the Brownian bridge representation for the generating function of $h(x,t)$ and $\tilde{h}(x,t)$ in the long time limit however suggests that any additional shift must vanish when $t\to\infty$.

The height fluctuations $\tilde{h}(x,t)$ should in principle be related to a solution of the KPZ equation as in (\ref{h[hKPZ] fluct}), but with singular initial condition akin to the much studied sharp wedge case \cite{SS2010.3,ACQ2011.1,CLDR2010.1,D2010.1} on the infinite line. An additional deterministic shift is expected compared to (\ref{h[hKPZ] fluct}). On the infinite line \cite{SS2010.2} or on an open interval \cite{CS2016.1}, this shift can be extracted from the comparison between the average $\langle\mathfrak{Z}_{\lambda,\ell}(x,t)\rangle$ of the Feynman-Kac solution of the stochastic heat equation obtained from KPZ by the Cole-Hopf transform $\mathfrak{Z}_{\lambda,\ell}(x,t)=\rme^{-2\lambda\mathfrak{h}_{\lambda,\ell}(x,t)}$, and G\"artner's microscopic Cole-Hopf transform for the exclusion process with partial asymmetry that verifies a closed equation. Unfortunately, the time evolution of G\"artner's transform does not seem to be straightforward in the case of periodic boundaries, and we were not able to state a precise connection to the KPZ equation with non-constant $\rho_{0}$.

Comparing (\ref{H[h] fluct}) and (\ref{H[h] finite}) for the two types of initial states of TASEP, we expect that $\tilde{h}(x,t)$ should be recovered from $h(x,t)$ in the limit where the initial amplitude $h_{0}$ of $h(x,t)$ is large. Writing explicitly the dependency on the initial condition as $h(x,t;h_{0})$ and $\tilde{h}(x,t;\mathcal{H}_{0})$, we conjecture that
\begin{equation}
\label{h(x,t;H/epsilon) -> htilde(x,t;H)}
h(x,t;\mathfrak{H}/\epsilon)\simeq\min[\mathfrak{H}]/\epsilon+\tilde{h}(x,t;\mathfrak{H})
\end{equation}
when $\epsilon\to0$. The conjecture is illustrated in figure \ref{fig parabola variance} for the variance of the height with parabolic initial condition. Besides, since the long time behaviour of Burgers' hydrodynamics (\ref{He(te) asymptotics}) on the Euler time scale $\tm\sim L$ depends only on the positions of the global minimum of $\mathcal{H}_{0}$, the statistics of $\tilde{h}(x,t)$ on the KPZ time scale $\tm\sim L^{3/2}$ should also be independent from $\mathcal{H}_{0}$ except for the positions $\kappa$ where its global minimum is reached. In the long time limit, these conjectures are supported by an explicit calculation of the limit $\epsilon\to0$ of Brownian bridge formulas in section \ref{section theta -> theta tilde}. Comparison with Bethe ansatz results for domain wall initial condition are also in agreement with (\ref{h(x,t;H/epsilon) -> htilde(x,t;H)}). In the following, we call \textit{sharp wedge} initial condition any finite initial density profile $\rho_{0}$ such that the corresponding height function $\mathcal{H}_{0}$ has a unique global minimum.
\end{subsubsection}

\end{subsection}

\begin{subsection}{Generating function of height fluctuations at late times}

Exact Bethe ansatz calculations \cite{P2016.1} for a few  specific initial conditions suggest that the moment generating function of the height fluctuations $h(x,t)$ and $\tilde{h}(x,t)$, defined respectively in (\ref{H[h] fluct}) and (\ref{H[h] finite}), is equal in the large $L$ limit to a dynamical partition function summing contributions from infinitely many configurations $r$ of particle-hole excitations at the edges of a Fermi sea. For general deterministic initial conditions $h_{0}$ or $\mathcal{H}_{0}$ defined above, we conjecture
\begin{eqnarray}
\label{GFh[theta_r] fluct}
&& \langle\rme^{sh(x,t)}\rangle=\sum_{r}\theta_{r}(s;h_{0})\,\rme^{\rmi p_{r}x+te_{r}(s)}\\
\label{GFh[theta_r] finite}
&& \langle\rme^{s\tilde{h}(x,t)}\rangle=\sum_{r}\tilde{\theta}_{r}(s;\mathcal{H}_{0})\,\rme^{\rmi p_{r}x+te_{r}(s)}\;.
\end{eqnarray}
For random initial condition, the coefficients $\theta_{r}$ and $\tilde{\theta}_{r}$ have to be averaged over the initial state. The total momentum $p_{r}$ is the sum of the momenta of the excitations. The function $e_{r}(s)$ in (\ref{GFh[theta_r] fluct}), (\ref{GFh[theta_r] finite}) has an explicit expression \cite{P2014.1} involving sums of momenta of excitations to the power $3/2$ (see \cite{GS1992.1,K1995.1,GM2005.1,dGE2005.1} for earlier results). At present, the coefficients $\theta_{r}(s;h_{0})$ in (\ref{GFh[theta_r] fluct}) are only known for flat ($h_{0}=0$) and stationary (Brownian) initial conditions whereas the coefficients $\tilde{\theta}_{r}(s;\mathcal{H}_{0})$ in (\ref{GFh[theta_r] finite}) are only known for sharp wedge initial condition \cite{P2016.1}.

In the long time limit, only the fully filled Fermi sea $r=0$ with no particle-hole excitation and zero momentum contributes, and one has
\begin{eqnarray}
\label{GFh[theta] fluct}
&& \langle\rme^{sh(x,t)}\rangle\underset{t\to\infty}{\simeq}\theta(s;h_{0})\,\rme^{te(s)}\\
\label{GFh[theta] finite}
&& \langle\rme^{s\tilde{h}(x,t)}\rangle\underset{t\to\infty}{\simeq}\tilde{\theta}(s;\mathcal{H}_{0})\,\rme^{te(s)}\;,
\end{eqnarray}
with $\theta(s;h_{0})=\theta_{0}(s;h_{0})$, $\tilde{\theta}(s;\mathcal{H}_{0})=\tilde{\theta}_{0}(s;\mathcal{H}_{0})$, $e(s)=e_{0}(s)$. The quantities $\theta(s;h_{0})$, $\tilde{\theta}(s;\mathcal{H}_{0})$ are written in (\ref{theta_r[psi_r] fluct}), (\ref{theta_r[psi_r] finite}) below as the large $L$ limit of an expression involving the elements of the stationary eigenstate of a deformed generator of TASEP. From the Perron-Frobenius theorem, all the components of this eigenvector are nonzero, at least for generic value of the deformation parameter, which ensures that for $r=0$ the right hand side in (\ref{theta_r[psi_r] fluct}), (\ref{theta_r[psi_r] finite}) is nonzero before taking the limit $L\to\infty$. We conjecture that this property still holds after the large $L$ limit, and that $\theta(s;h_{0})$, $\tilde{\theta}(s;\mathcal{H}_{0})$ are nonzero. This conjecture can also be seen as a consequence of the expected ergodicity of KPZ dynamics in finite volume, which implies that $\lim_{t\to\infty}t^{-1}\log\langle\rme^{sh(x,t)}\rangle$ may not depend on the initial condition. Depending on $h_{0}$, higher $\theta_{r}(s;h_{0})$ with $r\neq0$ may however vanish. This is for instance the case with flat initial condition, for which only eigenstates with momentum $p_{r}=0$ contribute to (\ref{GFh[theta_r] fluct}).

If we consider a dynamics which is also conditioned on the final state $h_{1}$ (respectively $\mathcal{H}_{1}$), we shall write $\theta(s;h_{0}\to h_{1})$ (resp. $\tilde{\theta}(s;\mathcal{H}_{0}\to\mathcal{H}_{1})$) for the coefficient in front of the exponential. We note that these functions $\theta$ appear as subleading prefactors to the dominant exponential behaviour. Similar corrections to large deviation asymptotics have been investigated for work fluctuations of Brownian particles \cite{PS2013.1} and heat transport in harmonic chains \cite{KSD2011.1}.
\end{subsection}

\begin{subsection}{Extremal values of independent Brownian bridges for $\theta(s;h_{0})$, $\tilde{\theta}(s;\mathcal{H}_{0})$ and $e(s)$}
\label{section theta bb}

One of our main results in this work is that we can relate the functions $\theta(s;h_{0})$, $\tilde{\theta}(s;\mathcal{H}_{0})$, with general initial condition $h_{0}(x)$, and $e(s)$, to statistical properties of standard Brownian bridges (see, for example, the expressions (\ref{theta[theta bb]}), (\ref{theta bb}) for $\theta(s;h_{0})$). Furthermore, we have also found a representation of these functions in terms of conditioned probabilities of non-intersecting Brownian bridges (see, e.g., the formula (\ref{theta bb non intersecting})).

\begin{subsubsection}{Relation to Brownian bridges.}
\hfill\break
The Wiener process $w(x)$, or standard Brownian motion, is a continuous random function with $w(0)=0$ whose increments $w(x+a)-w(x)$, $a>0$ are independent from $w(y)$, $y\leq x$ and have Gaussian distribution with mean $0$ and variance $a$. The standard Brownian bridge $b(x)$, $0\leq x\leq1$ (see figure \ref{fig bb}) is constructed from the Wiener process $w(x)$ by conditioning the latter on the event $w(1)=0$. Equivalently, the Brownian bridge may also be defined as $b(x)=w(x)-xw(1)$, which is convenient for simulations. The standard Brownian bridge $b(x)$ is a Gaussian process with mean $0$ and covariance $\langle b(x_{1})b(x_{2})\rangle_{b}=x_{1}(1-x_{2})$ for $0\leq x_{1}\leq x_{2}\leq1$.

\begin{figure}
  \begin{center}
    \includegraphics[width=100mm]{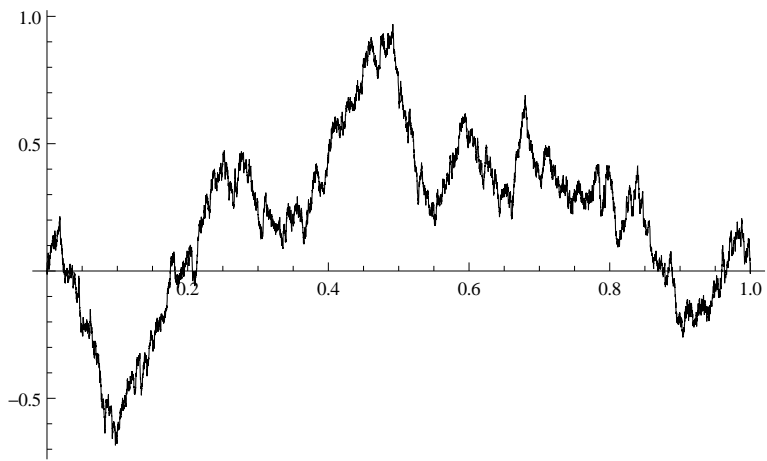}
  \end{center}
  \caption{Typical realization $b(x)$ of the standard Brownian bridge plotted as a function of $x$.}
  \label{fig bb}
\end{figure}

In the long time limit, the height $h(x,t)$ becomes Brownian, in the sense that $h(x,t)-h(0,t)$ has the same law as a standard Brownian bridge $b(x)$ \cite{HHZ1995.1}. This result does not say anything, however, about the correlations between $h(0,t)$ and $b(x)$. Using the matrix product representation obtained by Lazarescu and Mallick in \cite{LM2011.1,L2013.1} for the dominant eigenstate of TASEP, we derive in section \ref{section matrix product} the perturbative expansion near $s=0$ to arbitrary order $n\in\mathbb{N}$
\begin{equation}
\label{theta[theta bb]}
\theta(s;h_{0})=\theta_{\bb,n}(s;h_{0})+\mathcal{O}(s^{n+1})\;,
\end{equation}
with
\begin{equation}
\label{theta bb}
\fl\hspace{5mm}
\boxed{
\theta_{\bb,n}(s;h_{0})=\frac{\langle\rme^{-s\sum_{j=1}^{n}\max[b_{j}-b_{j-1}]}\rangle_{b_{0},\ldots,b_{n}}\langle\rme^{-s\max[b_{1}-h_{0}]-s\sum_{j=2}^{n}\max[b_{j}-b_{j-1}]}\rangle_{b_{1},\ldots,b_{n}}}{\langle\rme^{-s\sum_{j=1}^{2n}\max[b_{j}-b_{j-1}]}\rangle_{b_{0},\ldots,b_{2n}}}
}\;.
\end{equation}
Here, $\max[f]$ denotes the maximum of a function $f(x)$ in the interval $x\in[0,1]$. Besides, expressions of the form $\langle\ldots\rangle_{b_{0},b_{1},\ldots}$ always indicate expectation values computed with respect to independent standard Brownian bridges $b_{j}$.

More generally, when the dynamics is conditioned on the final height profile $h(x,t)-h(0,t)=h_{1}(x)$, the same approach gives for $\theta_{\bb,n}(s;h_{0}\to h_{1})$ a formula obtained by replacing in the numerator of (\ref{theta bb}) the factor $\langle\rme^{-s\sum_{j=1}^{n}\max[b_{j}-b_{j-1}]}\rangle_{b_{0},\ldots,b_{n}}$ by $\langle\rme^{-s\sum_{j=2}^{n}\max[b_{j}-b_{j-1}]-s\max[h_{1}-b_{n}]}\rangle_{b_{1},\ldots,b_{n}}$. Note that the fact that the stationary state corresponds to a random height profile with the statistics of a standard Brownian bridge amounts to $\theta_{\bb,n}(s;h_{0})=\langle\theta_{\bb,n}(s;h_{0}\to b)\rangle_{b}$. Additionally, one has $\langle\theta_{\bb,n}(s;h_{0}\to b)\,\theta_{\bb,n}(s;b\to h_{1})\rangle_{b}=\theta_{\bb,n}(s;h_{0}\to h_{1})$, which is consistent with the insertion of an intermediate time in $\langle\rme^{sh(x,t)}\rangle$.

The large $L$ limit of the eigenvalue equation for the dominant eigenstate of TASEP is given in section \ref{section bb master equation} in terms of Brownian averages. One finds the perturbative expansion near $s=0$ to arbitrary order $n\in\mathbb{N}$
\begin{equation}
\label{e[f bb]}
\frac{e(s)}{s}+\frac{s^{2}}{3}=f_{\bb,n}(s;h_{0})+\mathcal{O}(s^{n})
\end{equation}
with
\begin{equation}
\label{f bb}
f_{\bb,n}(s;h_{0})=-\frac{\langle b_{n}'(0^{+})\,b_{n}'(1^{-})\,\rme^{-s\max[b_{1}-h_{0}]-s\sum_{j=2}^{n}\max[b_{j}-b_{j-1}]}\rangle_{b_{1},\ldots,b_{n}}}{\langle\rme^{-s\max[b_{1}-h_{0}]-s\sum_{j=2}^{n}\max[b_{j}-b_{j-1}]}\rangle_{b_{1},\ldots,b_{n}}}\;.\nonumber
\end{equation}
The derivatives $b_{n}'(0^{+})$ and $b_{n}'(1^{-})$ in (\ref{f bb}) are to be understood as the limits $b_{n}'(0^{+})=\lim_{M\to\infty}Mb_{n}(1/M)$, $b_{n}'(1^{-})=-\lim_{M\to\infty}Mb_{n}(1-1/M)$. The function $h_{0}$ in (\ref{f bb}) is an arbitrary, regular enough continuous function with $h_{0}(0)=h_{0}(1)=0$. As an eigenvalue equation, (\ref{f bb}) must be independent from the overlaps of the corresponding eigenvector. Thus, the perturbative expansion of $f_{\bb,n}(s;h_{0})$ up to order $s^{n-1}$ must a priori not depend of $h_{0}$. This can be checked directly by noting that (\ref{f bb}) generates no chain of the form $\max[b_{1}-h_{0}]^{r_{1}}\max[b_{2}-b_{1}]^{r_{2}}\ldots\max[b_{n}-b_{n-1}]^{r_{n}}$, $r_{1},\ldots,r_{n}\geq1$ linking $h_{0}$ and $b_{n}$ at that order. 

\vskip 0.2cm
\noindent
\textbf{Remark:} Translation invariance.
The current $Q_{i}(\tm)$ of TASEP between sites $i$ and $i+1$ with initial configuration $\mathcal{C}_{0}$ is unchanged if one translates both $i$ and all the particles in the initial state by the same distance. Using $Q_{i}(\tm)=H_{i}(\tm)-H_{i}(0)$ and the large $L$ asymptotics (\ref{H[h] fluct}), this implies that $h(x,t)-h_{0}(x)$ is invariant under a simultaneous translation by an arbitrary distance $a$ of the position $x$ and the initial density profile $\sigma_{0}$. In the long time limit, this is equivalent to
\begin{equation}
\theta(s;h_{0})=\rme^{sh_{0}(-a)}\,\theta(s;h_{0}(\,\cdot\,-a)-h_{0}(-a))
\end{equation}
for any periodic function $h_{0}$ with period $1$.

At first order in $s$, the identity above reduces from (\ref{theta bb}) to an average translation invariance property of the Brownian bridge, that $\langle\max_{0\leq x\leq1}[b(x)-h_{0}(x-a)]\rangle_{b}$ is independent of $a$ for any continuous function $h_{0}$ of period $1$. The latter property can be proved by writing $b(x+a)-h_{0}(x)=b(x)-h_{0}(x)+(b(x+a)-b(x))$, where $b(x+a)-b(x)=w(x+a)-w(x)-aw(1)$ is equal in law to $\hat{w}(a)-aw(1)$ with $w$ and $\hat{w}$ Wiener processes of mean zero.

We emphasize that translation invariance holds for $\theta(s;h_{0})$ but not for $\theta_{\bb,n}(s;h_{0})$ with finite $n$ and $s$: only the perturbative expansion up to order $n$ in $s$ has to be invariant when $h_{0}(x)$ is replaced by $h_{0}(x-a)-h_{0}(-a)$.
\end{subsubsection}

\begin{subsubsection}{Brownian bridge for $\tilde{\theta}(s;\mathcal{H}_{0})$.}
\label{section theta -> theta tilde}
\hfill\break
From (\ref{h(x,t;H/epsilon) -> htilde(x,t;H)}), $\theta(s;\mathfrak{H}/\epsilon)$ and $\tilde{\theta}(s;\mathfrak{H})$ are related when $\epsilon\to0$. Furthermore, the discussion at the end of section \ref{section Burgers} about hydrodynamics at long times on the Euler scale indicates that $\tilde{\theta}(s;\mathfrak{H})$ should not depend of $\mathfrak{H}$ except for the locations at which its global minimum is reached.

The situation is particularly simple for sharp wedge initial condition
 (that we shall denote by the acronym $\text{sw}$ in the following) where the global minimum is reached only once, at $\kappa\in(0,1]$. For any $x\in(0,1]$, $x\neq\kappa$ and any realization $b$ of the standard Brownian bridge the inequality $b(\kappa)-\epsilon^{-1}\mathfrak{H}(\kappa)>b(x)-\epsilon^{-1}\mathfrak{H}(x)$ holds for small enough epsilon. Thus, the random variable $\max[b-\epsilon^{-1}\mathfrak{H}]+\epsilon^{-1}\min[\mathfrak{H}]\to b(\kappa)$ almost surely when $\epsilon\to0$. Using translation invariance, one can as well take $\kappa=0$ so that $b(\kappa)=0$, and (\ref{theta bb}) leads to
\begin{equation}
\label{theta(s;H/epsilon) -> thetatilde(s;H)}
\lim_{\epsilon\to0}\rme^{-s\min[\mathfrak{H}]/\epsilon}\theta(s;\mathfrak{H}/\epsilon)=\tilde{\theta}(s;\text{sw})\;,
\end{equation}
where
\begin{equation}
\label{theta sw[theta bb sw]}
\tilde{\theta}(s;\text{sw})=\tilde{\theta}_{\bb,n}(s;\text{sw})+\mathcal{O}(s^{n+1})
\end{equation}
with
\begin{equation}
\label{theta bb sw}
\tilde{\theta}_{\bb,n}(s;\text{sw})=\frac{\langle\rme^{-s\sum_{j=1}^{n}\max[b_{j}-b_{j-1}]}\rangle_{b_{0},\ldots,b_{n}}\langle\rme^{-s\sum_{j=2}^{n}\max[b_{j}-b_{j-1}]}\rangle_{b_{1},\ldots,b_{n}}}{\langle\rme^{-s\sum_{j=1}^{2n}\max[b_{j}-b_{j-1}]}\rangle_{b_{0},\ldots,b_{2n}}}\;.
\end{equation}
This result is checked in section \ref{section dw} up to order $s^{3}$ for domain wall initial condition by comparison with the exact Bethe ansatz result.

When the global minimum of $\mathfrak{H}$ is reached several times, say at discrete positions $\kappa_{1}$, $\kappa_{2}$, $\ldots$ the procedure above replaces the factor $\rme^{-s\max[b_{1}-\epsilon^{-1}\mathfrak{H}]}$ in (\ref{theta bb}) by $\rme^{s\min[\mathfrak{H}]/\epsilon}\,\rme^{-s\max(b_{1}(\kappa_{1}),b_{1}(\kappa_{2}),\ldots)}$.
 Therefore, the result depends explicitly on the distances between the positions $\kappa_{j}$ after using translation invariance. In particular, the mean value of the height $\langle h(x,t)\rangle$ in the long time limit is shifted by the amount $-\langle\max(b(\kappa_{1}),b(\kappa_{2}),\ldots)\rangle_{b}$ as compared to the sharp wedge case.
\end{subsubsection}

\end{subsection}

\begin{subsection}{First cumulants of the height}
\label{section cumulants}
The function $s\mapsto\log\langle\rme^{sh(x,t)}\rangle$ is the cumulant generating function of the height. At late times, (\ref{GFh[theta] fluct}) implies that the cumulant generating function is equal to $te(s)+\log\theta(s;h_{0})$ up to exponentially small corrections. The Legendre transform $g$ of $e$ is the large deviation function of the height in the stationary state, i.e., the probability density of the height behaves as $\P(h(x,t)=tu)\sim\rme^{-tg(u)}$ for large $t$.

Because of translation invariance, the cumulants of the height are independent of the position $x$ in the long time limit. The leading term $e(s)$ is also independent of the initial condition. Its expression,
 see (\ref{e[chi]}) below, was first obtained by Derrida and Lebowitz \cite{DL1998.1} for TASEP, see also \cite{DA1999.1}, and recovered by Brunet and Derrida \cite{BD2000.1} for the continuum directed polymer in a random medium, which also belongs to KPZ universality.

\begin{figure}
  \begin{center}
    \begin{tabular}{cc}
      \includegraphics[width=75mm]{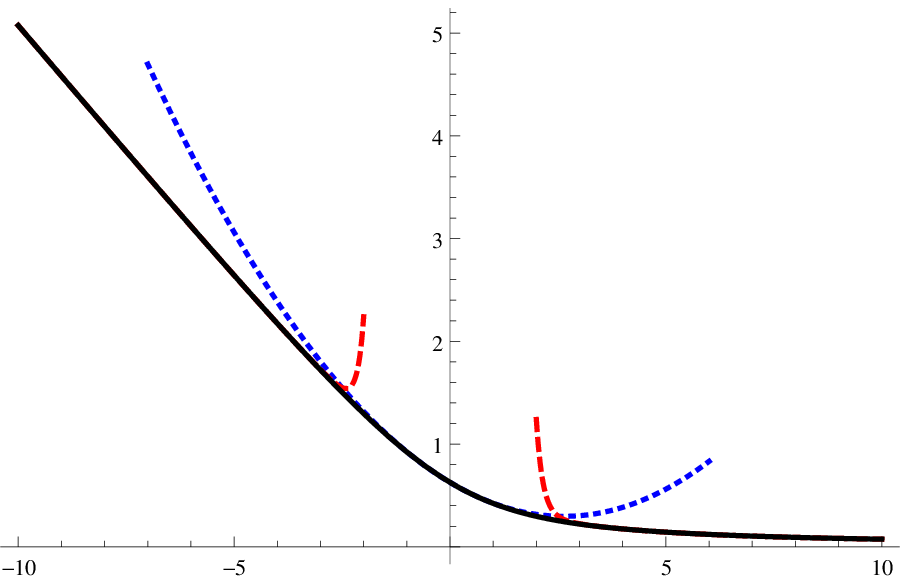}
      &
      \includegraphics[width=75mm]{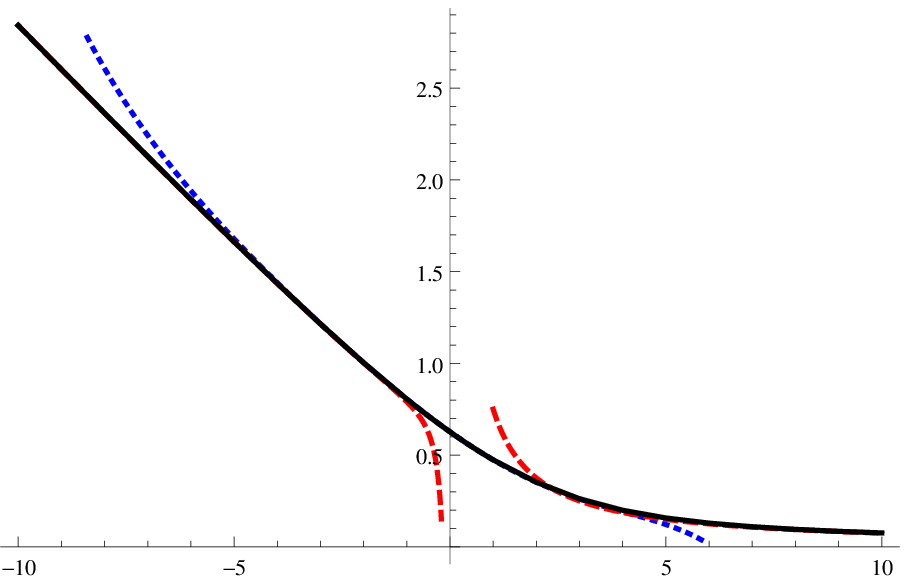}
    \end{tabular}
  \end{center}
  \caption{{Plot of $\langle\max_{0\leq x\leq1}[b(x)-h(x)]\rangle_{b}$ for $h(x)=cx\openone_{\{0\leq x<1/2\}}+c(1-x)\openone_{\{1/2\leq x<1\}}$ (left) and $h(x)=cx(1-x)$ (right) as a function of $c$. The average is computed with respect to the standard Brownian bridge $b$. On the left, the black curve is the exact formula (\ref{max[b-h] piecewise linear symmetric}), the dashed red curves are the asymptotics $c\to-\infty$ (\ref{max[b-h] piecewise linear c<<-1}) and $c\to+\infty$ (\ref{max[b-h] piecewise linear c>>1}) with $a=1/2$ and summation up to $k=6$, and the dotted blue curve is the small $c$ expansion (\ref{max[b-h] piecewise linear small c}) with $a=1/2$. On the right, the black curve comes from high precision Bethe ansatz numerics (and matches perfectly (\ref{max[b-h] parabola int lambda}) for $c<0$), the dashed red curves are the asymptotics $c\to-\infty$ (\ref{max[b-h] parabola c<<-1}) and $c\to+\infty$ (\ref{max[b-h] parabola c>>1}), and the dotted blue curve is the small $c$ expansion (\ref{max[b-h] parabola small c}).}}
  \label{fig piecewise and parabola}
\end{figure}

\begin{subsubsection}{Average height.}
\label{section <h>}
\hfill\break 
Using the fact that $\langle\max[b_{1}-b_{0}]\rangle_{b_{0},b_{1}}=\sqrt{\pi}/2$ (this expectation value is a special case of equation (\ref{<max[b1-b0]^r>}) below), we deduce from (\ref{theta bb}) that the mean value of the height in the case of an initial condition of the form (\ref{rho0 fluct}) is equal in the long time limit to
\begin{equation}
\langle h(x,t)\rangle\simeq t+\frac{\sqrt{\pi}}{2}-\langle\max[b-h_{0}]\rangle_{b}
\end{equation}
\textit{up to exponentially small corrections.} The leading term $t$ comes from $e'(0)=1$, which is a consequence of the exact expression (\ref{e[chi]}) below.

In order to get a closed expression for the average, we need to calculate the average $\langle\max[b-h_{0}]\rangle_{b}$ over Brownian bridges with the initial condition $h_{0}$. We shall discuss now some special cases.

In the case of a flat initial condition $h_{0}(x)=0$, we obtain $\langle h(x,t)\rangle\simeq t+\frac{\sqrt{\pi}}{2}-\frac{\sqrt{2\pi}}{4}$. For a stationary initial condition $h_{0}(x)=b(x)$ with $b$ a standard Brownian bridge, the average cancels out the $\sqrt{\pi}/{2}$ term and $\lim_{t\to\infty}\langle h(x,t)\rangle-t=0$. Special cases with piecewise linear and parabolic initial conditions are considered in figure \ref{fig piecewise and parabola}.

When $h_{0}$ is of small amplitude, the following expansion is derived in section \ref{section <max[b-h]> small h}:
\begin{equation}
\label{max[b-h] small h}
\fl\hspace{1mm}
\langle\max[b-\epsilon h_{0}]\rangle_{b}=\frac{\sqrt{2\pi}}{4}-\epsilon\!\int_{0}^{1}\!\rmd x\,h_{0}(x)-\epsilon^{2}\sqrt{2\pi}\,\sum_{k\in\mathbb{Z}}a_{k}a_{-k}(-1)^{k}k\pi J_{1}(k\pi)+\mathcal{O}(\epsilon^{3})\,,
\end{equation}
where the $a_{k}$ are the Fourier coefficients of $h_{0}$, $h_{0}(x)=\sum_{k\in\mathbb{Z}}a_{k}\rme^{2\rmi\pi kx}$ and $J$ is the Bessel function of the first kind.

On the other hand, when $h_{0}$ is of large amplitude, we recover from (\ref{h(x,t;H/epsilon) -> htilde(x,t;H)}) the average $\langle\tilde{h}(x,t)\rangle$ corresponding to a finite initial density profile $\rho_{0}$. In particular, if the minimum of the initial height is reached only once, the sharp wedge case is recovered in the limit of large amplitude, and (\ref{theta bb sw}) implies $\langle\tilde{h}(x,t)\rangle\simeq t+\frac{\sqrt{\pi}}{2}$. More precisely, the asymptotics for $h_{0}$ of large amplitude is dominated by the behaviour of $h_{0}$ around its global minimum. If the minimum is reached only once, at $\kappa$ with $h_{0}(x)\simeq\min[h_{0}]+a|x-\kappa|^{\nu}$, the scaling properties of the Brownian motion give at leading orders in $\epsilon$ the asymptotics $\langle\max_{0\leq x\leq1}(b(x)-h_{0}(x)/\epsilon)\rangle_{b}\simeq-\epsilon^{-1}\min[h_{0}]+(\epsilon/a)^{\frac{1}{2\nu-1}}\langle\max_{x\in\mathbb{R}}(w(x)-|x|^{\nu})\rangle_{w}$ when $\nu>1/2$, with $w$ a two-sided Wiener process. For $0\leq\nu\leq1/2$, the correction to the leading term $-\epsilon^{-1}\min[h_{0}]$ becomes instead exponentially small when $\epsilon\to0$. When $\nu=1$, the constant $\langle\max_{x\in\mathbb{R}}(w(x)-|x|)\rangle_{w}=3/4$ follows from $\P(\max_{x\in\mathbb{R}}(w(x)-|x|)<z)=(1-\rme^{-2z})^{2}$ for $z>0$, see section \ref{section piecewise linear}. When $\nu=2$, the constant $\langle\max_{x\in\mathbb{R}}(w(x)-x^{2})\rangle_{w}=\Xi/2^{2/3}$ can be computed explicitly in terms of the Airy function as (\ref{Xi parabola int lambda}), (\ref{Xi parabola sum j}) using exact results by Groeneboom \cite{G1989.1} about the Wiener process absorbed by a parabola, see also \cite{S1988.2,DS1985.1,JLML2010.1,KR2010.1}.

In the particular case when $h_{0}$ is piecewise linear, $h_{0}(x)=cx\openone_{\{0\leq x<1/2\}}+c(1-x)\openone_{\{1/2\leq x\leq1\}}$, $0\leq x\leq1$, the Brownian bridge average $\langle\max[b-h_{0}]\rangle_{b}$ can be computed exactly for finite $c$ using standard techniques, see section \ref{section piecewise linear}. For large $|c|$, one finds the asymptotics $\langle\max[b-h_{0}]\rangle_{b}\simeq\frac{3}{4c}$ for $c>0$ and $\langle\max[b-h_{0}]\rangle_{b}\simeq\frac{|c|}{2}+\frac{3}{4|c|}$ for $c<0$. This is consistent with the discussion in the previous paragraph since the behaviour of periodized $h_{0}$ around its minimum is $h_{0}(x)\simeq c|x|$ for $c>0$ and $h_{0}(x)\simeq-\frac{|c|}{2}-c|x-1/2|$ for $c<0$.

In the quadratic case $h_{0}(x)=cx(1-x)$, exact results \cite{G1989.1} for the Wiener process absorbed by a parabola give the explicit non-perturbative expressions (\ref{max[b-h] parabola int lambda}) or (\ref{max[b-h] parabola sum j}) for $\langle\max[b-h_{0}]\rangle_{b}$. For large $|c|$, $c<0$, one has in particular $\langle\max[b(x)-h_{0}]\rangle_{b}\simeq-\frac{c}{4}+\frac{\Xi}{|4c|^{1/3}}+\frac{1}{4c}$ up to exponentially small corrections, with the constant $\Xi\approx1.25512$ defined in terms of the Airy function in (\ref{Xi parabola int lambda}) or (\ref{Xi parabola sum j}). The first two terms are in agreement with the expansion for $h_{0}$ of large amplitude above since $cx(1-x)\simeq-\frac{|c|}{4}+|c|(x-1/2)^{2}$ when $x\to1/2$. For large positive $c$, one has instead $\langle\max[b-h_{0}]\rangle_{b}\simeq\frac{3}{4c}$, which is also consistent with the discussion above since $cx(1-x)\simeq c|x|$ when $x\to0$ modulo $1$.
\end{subsubsection}

\begin{figure}
  \begin{center}
    \includegraphics[width=75mm]{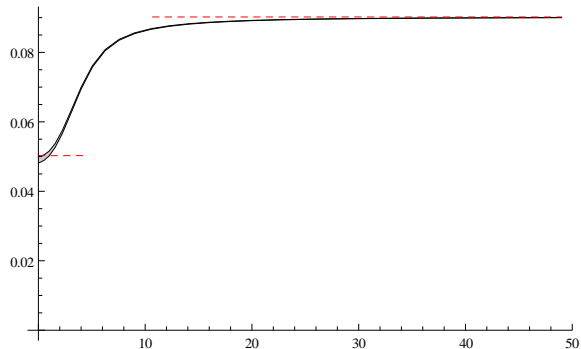}
  \end{center}
  \caption{Plot of the variance of the height $\langle h(x,t)^{2}\rangle-\langle h(x,t)\rangle^{2}$, minus its asymptotic value $\frac{\sqrt{\pi}}{2}\,t$ in the long time limit, as a function of the amplitude $c$ of the initial height $h_{0}$ in the parabolic case $h_{0}(x)=cx(1-x)$. The variance is computed numerically from (\ref{variance h}) by averaging Brownian bridges discretized uniformly with $10^6$ points. The upper and lower curves correspond to the statistical average plus and minus the standard deviation for an averaging over $10^6$ Brownian bridges. The red, dashed asymptotes at small and large amplitude $c$ are respectively given by (\ref{variance h flat}) and (\ref{variance h sw}).}
  \label{fig parabola variance}
\end{figure}

\begin{subsubsection}{Variance of the height.}
\label{section variance h}
\hfill\break 
Using (\ref{theta bb}) again and the explicit Brownian averages (\ref{<max[b]^r>}), (\ref{<max[b1-b0]^r>}), (\ref{<max[b1-b0]max[b2-b1]>}) and (\ref{<max[b0]max[b1-b0]>}), we obtain that the variance of the height for the case of an initial condition of the form (\ref{rho0 fluct}) is equal in the long time limit to
\begin{eqnarray}
\label{variance h}
&&\fl\hspace{2mm} \langle h(x,t)^{2}\rangle-\langle h(x,t)\rangle^{2}
\simeq\frac{\sqrt{\pi}}{2}\,t+1+\frac{5\pi}{4}-\frac{8\pi}{3\sqrt{3}}-\sqrt{\pi}\,\langle\max[b-h_{0}]\rangle_{b}\\
&& +\langle\max[b-h_{0}]^{2}\rangle_{b}-\langle\max[b-h_{0}]\rangle_{b}^{2}+2\langle\max[b_{1}-h_{0}]\max[b_{2}-b_{1}]\rangle_{b_{1},b_{2}}\nonumber
\end{eqnarray}
up to exponentially small corrections. The leading term in $t$, equal to $e''(0)t$, follows from the exact expression (\ref{e[chi]}) below.

For a flat initial condition $h_{0}(x)=0$, we obtain in particular
\begin{equation}
\label{variance h flat}
\langle h(x,t)^{2}\rangle-\langle h(x,t)\rangle^{2}\simeq\frac{\sqrt{\pi}}{2}\,t+\frac{1}{2}+\Big(\frac{7}{4}-\frac{1}{2\sqrt{2}}-\frac{8}{3\sqrt{3}}\Big)\pi
\end{equation}
For a stationary initial condition $h_{0}(x)=b(x)$ with $b$ a standard Brownian bridge, we have 
\begin{equation}
\label{variance h stat}
\langle h(x,t)^{2}\rangle-\langle h(x,t)\rangle^{2}\simeq\frac{\sqrt{\pi}}{2}\,t+1+\Big(\frac{1}{2}-\frac{4}{3\sqrt{3}}\Big)\pi\;.
\end{equation}

Similarly, using (\ref{theta bb sw}), we deduce that the variance of the height in the sharp wedge case is equal in the long time limit to
\begin{equation}
\label{variance h sw}
\langle\tilde{h}(x,t)^{2}\rangle-\langle\tilde{h}(x,t)\rangle^{2}\simeq\frac{\sqrt{\pi}}{2}\,t+1+\Big(\frac{5}{4}-\frac{8}{3\sqrt{3}}\Big)\pi\;.
\end{equation}
The crossover between flat and sharp wedge initial conditions is plotted in figure \ref{fig parabola variance} for the example of a parabolic interface.
\end{subsubsection}

\begin{subsubsection}{Multiple point correlations.}
\hfill\break
The Brownian representation for the one-point generating function $\langle\rme^{sh(x,t)}\rangle$ in the long time limit, discussed in the previous sections, can readily be generalized to multiple-point correlations. Introducing a function $s(x)$, $0\leq x\leq1$, one has
\begin{equation}
\langle\rme^{\int_{0}^{1}\rmd x\,s(x)h(x,t)}\rangle\simeq\Theta[s;h_{0}]\,\rme^{te(\overline{s})}
\end{equation}
with $\overline{s}=\int_{0}^{1}\rmd x\,s(x)$. The prefactor $\Theta[s;h_{0}]$ is now a functional
 of the function $s(x)$ (and it must not be confused with $\theta(s;h_{0})$ with $s$ scalar considered in the previous sections) verifies $\Theta[s;h_{0}]=\Theta_{\bb,n}[s;h_{0}]+\mathcal{O}(s^{n+1})$ where $\Theta_{\bb,n}[s;h_{0}]$ can be expressed in terms of standard Brownian bridges as
\begin{eqnarray}
\label{Theta bb}
&& \Theta_{\bb,n}[s;h_{0}]=\langle\rme^{\int_{0}^{1}\rmd x\,s(x)b_{n}(x)-\overline{s}\sum_{j=1}^{n}\max[b_{j}-b_{j-1}]}\rangle_{b_{0},\ldots,b_{n}}\\
&&\hspace{23mm} \times\,\frac{\langle\rme^{-\overline{s}\max[b_{1}-h_{0}]-\overline{s}\sum_{j=2}^{n}\max[b_{j}-b_{j-1}]}\rangle_{b_{1},\ldots,b_{n}}}{\langle\rme^{-\overline{s}\sum_{j=1}^{2n}\max[b_{j}-b_{j-1}]}\rangle_{b_{0},\ldots,b_{2n}}}\;.\nonumber
\end{eqnarray}
Note that the special case $s(x)=-s\,\delta'(x-x_{0})$, with $\overline{s}=0$, leads to $\langle\rme^{s\,\partial_{x}h(x,t)}\rangle\to\langle\rme^{s\,b'(x)}\rangle_{b}$ when $t\to\infty$, which is consistent with the fact that the stationary state of the KPZ equation, up to a global shift removing the average drift, is Brownian (i.e. $h(x,t)-h(0,t)$ has the same statistics as a standard Brownian bridge $b(x)$ correlated to $h(0,t)$ but decoupled from the initial condition $h_{0}$, as indicated by (\ref{Theta bb})).

The Family-Vicsek scaling function \cite{FV1985.1} $f_{\text{FV}}(t)=\langle(\int_{0}^{1}\rmd x\,h^{2}(x,t))-(\int_{0}^{1}\rmd x\,h(x,t))^{2}\rangle$, which characterizes the width of the interface, is another example involving multiple-point statistics of the height. Since $f_{\text{FV}}(t)=\langle\int_{0}^{1}\rmd x\,(h(x,t)-\overline{h}(t))^{2}\rangle$ with $\overline{h}(t)=\int_{0}^{1}\rmd x\,h(x,t)$ and $h(x,t)-\overline{h}(t)\to b(x)-\int_{0}^{1}\rmd y\,b(y)$ when $t\to\infty$ with $b$ a standard Brownian bridge from the previous paragraph, one has $f_{\text{FV}}(\infty)=\langle(\int_{0}^{1}\rmd x\,b^{2}(x))-(\int_{0}^{1}\rmd x\,b(x))^{2}\rangle_{b}$, and the covariance of the standard Brownian bridge implies $f_{\text{FV}}(\infty)=1/12$. An alternative derivation, which follows more directly from (\ref{Theta bb}), consists in using translation invariance of the moments of $h(x,t)$ to write
\begin{eqnarray}
&& f_{\text{FV}}(\infty)=\lim_{t\to\infty}\langle h(x,t)^{2}\rangle-\Big\langle\Big(\int_{0}^{1}\rmd x\,h(x,t)\Big)^{2}\Big\rangle\\
&&\hspace{14mm} =\partial_{s}^{2}\Big(\langle\rme^{sh(x,t)}\rangle-\langle\rme^{s\int_{0}^{1}\rmd x\,h(x,t)}\rangle\Big)_{|s\to0}\;.\nonumber
\end{eqnarray}
From (\ref{theta bb}), and (\ref{Theta bb}) with $s(x)$ constant, one finds $f_{\text{FV}}(\infty)=-\langle(\int_{0}^{1}\rmd x\,b(x))^{2}\rangle_{b}+2\langle\max[b_{2}-b_{1}]\int_{0}^{1}\rmd x\,b_{2}(x)\rangle_{b_{1},b_{2}}$. Comparison with $f_{\text{FV}}(\infty)=1/12$ implies the statistical identity
\begin{equation}
\langle\max[b_{2}-b_{1}]\int_{0}^{1}\rmd x\,b_{2}(x)\rangle_{b_{1},b_{2}}=\frac{1}{12}\;,
\end{equation}
that we have checked numerically.
\end{subsubsection}

\begin{subsubsection}{Multiple time correlations.}
\hfill\break
Correlations between multiple times $t_{1},t_{2},\ldots$ taken \textit{far apart} can also be computed using the fact that the system reaches stationarity at the intermediate times, i.e. each of the $h(x,t_{i})-h(0,t_{i})$ are independent Brownian bridges in $x$.

For simplicity, we consider only the two-time correlation $\langle\rme^{s_{1}h(0,t_{1})+s_{2}h(x,t_{2})}\rangle$ with $0\ll t_{1}\ll t_{2}$. Writing $\rme^{s_{1}h(0,t_{1})+s_{2}h(x,t_{2})}=\rme^{(s_{1}+s_{2})h(0,t_{1})}\,\rme^{s_{2}(h(x,t_{2})-h(0,t_{1}))}$, the first factor $\rme^{(s_{1}+s_{2})h(0,t_{1})}$ depends only on the evolution between time $0$ and time $t_{1}$, while in the second factor, $h(x,t_{2})-h(0,t_{1})$ has the same distribution as the KPZ height at position $x$ and time $t_{2}-t_{1}$ with Brownian bridge initial condition $b(x')=h(x',t_{1})-h(0,t_{1})$. This leads to
\begin{equation}
\langle\rme^{s_{1}h(0,t_{1})+s_{2}h(x,t_{2})}\rangle\simeq\langle\theta(s_{1}+s_{2};h_{0}\to b)\theta(s_{2};b)\rangle_{b}\,\rme^{t_{1}e(s_{1}+s_{2})+(t_{2}-t_{1})e(s_{2})}\;.
\end{equation}
The coefficient $\theta(s_{1}+s_{2};h_{0}\to b)$, corresponding to an evolution conditioned on the final state $b$, has a perturbative expansion for small $s_{1}+s_{2}$ in terms of Brownian bridges given below (\ref{theta bb}).
\end{subsubsection}

\end{subsection}

\begin{subsection}{Conditional probabilities of non-intersecting Brownian bridges}
\label{section non intersecting}
The expressions of the previous section involving extremal values of independent standard Brownian bridges can be conveniently rewritten in terms of conditional probabilities of \textit{non-intersecting} Brownian bridges.

We introduce Brownian bridges $\mathfrak{b}_{j}^{s}$, $j\in\mathbb{Z}$ depending on a parameter $s>0$ by
\begin{eqnarray}
\begin{array}{ll}
\label{bb non intersecting}
\mathfrak{b}_{j}^{s}(x)=b_{j}(x)-\sum_{k=1}^{|j|}z_{-k} & j<0\\
\mathfrak{b}_{0}^{s}(x)=b_{0}(x) & j=0\\
\mathfrak{b}_{j}^{s}(x)=b_{j}(x)-\sum_{k=1}^{j}z_{k} & j>0
\end{array}\;,
\end{eqnarray}
see figure \ref{fig bb non intersecting}. The $b_{j}$, $j\in\mathbb{Z}$ are independent standard Brownian bridges (with endpoints $b_{j}(0)=b_{j}(1)=0$) and the $z_{k}$, $k\in\mathbb{Z}$ are independent (from each other and from the $b_{j}$'s) exponentially distributed random variables with parameter $s$ of density $s\,\rme^{-sz}$. Equivalently, the endpoints $\mathfrak{b}_{j}^{s}(0)=\mathfrak{b}_{j}^{s}(1)$ are consecutive events of a Poisson point process with rate $s$ conditioned on an event at the origin. The bridges $\mathfrak{b}_{j}^{s}$ are not independent because their endpoints are correlated. For any $j<0<k$, however, $\mathfrak{b}_{j}^{s}$, $\mathfrak{b}_{0}^{s}$ and $\mathfrak{b}_{k}^{s}$ are independent.

\begin{figure}
  \begin{center}
    \includegraphics[width=100mm]{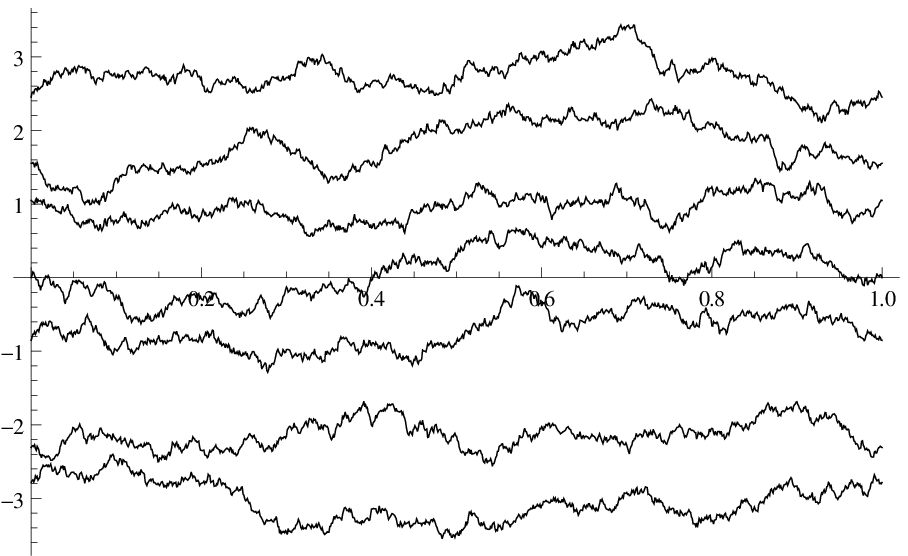}
    \begin{picture}(0,0)
      \put(-1,50){\small$\mathfrak{b}_{3}^{s}(x)$}
      \put(-1,30){\small$\mathfrak{b}_{0}^{s}(x)$}
      \put(-1,6){\small$\mathfrak{b}_{-3}^{s}(x)$}
    \end{picture}
  \end{center}
  \caption{Plot for $j=-3,\ldots,3$ and $s=2$ of a typical realization of the Brownian bridges $\mathfrak{b}_{j}^{s}(x)$ defined in (\ref{bb non intersecting}) and conditioned never  to  intersect.}
  \label{fig bb non intersecting}
\end{figure}

The Brownian bridge expectation value $\langle\rme^{-s\max[b_{1}-h_{0}]-s\sum_{j=2}^{n}\max[b_{j}-b_{j-1}]}\rangle_{b_{1},\ldots,b_{n}}$ in (\ref{theta bb}) is equal to $\int_{0}^{\infty}\rmd z_{1}\ldots\rmd z_{n}\,\rme^{-s(z_{1}+\ldots+z_{n})}\partial_{z_{1}}\ldots\partial_{z_{n}}F(z_{1},\ldots,z_{n};h_{0})$, with $F$ given by $F(z_{1},\ldots,z_{n};h_{0})=\P(b_{1}-h_{0}<z_{1},b_{2}-b_{1}<z_{2},\ldots,b_{n}-b_{n-1}<z_{n})$. When $s>0$, the derivatives can be eliminated using partial integration. Using $F(\ldots,0,\ldots)=0$, one obtains 
\begin{eqnarray}
\label{GF -> non intersecting bb}
&& \langle\rme^{-s\max[b_{1}-h_{0}]-s\sum_{j=2}^{n}\max[b_{j}-b_{j-1}]}\rangle_{b_{1},\ldots,b_{n}}\\
&&\hspace{25mm} =\int_{0}^{\infty}\rmd z_{1}\ldots\rmd z_{n}\,s^{n}\rme^{-s(z_{1}+\ldots+z_{n})}F(z_{1},\ldots,z_{n};h_{0})\;.\nonumber
\end{eqnarray}
This formula can be interpreted as the probability that $h_{0}$, $\mathfrak{b}_{-1}^{s}$, $\ldots$, $\mathfrak{b}_{-n}^{s}$ are non-intersecting:
\begin{equation}
\langle\rme^{-s\max[b_{1}-h_{0}]-s\sum_{j=2}^{n}\max[b_{j}-b_{j-1}]}\rangle_{b_{1},\ldots,b_{n}}=\P(\mathfrak{b}_{-n}^{s}<\ldots<\mathfrak{b}_{-1}^{s}<h_{0})\;.
\end{equation}
Therefore, the coefficient $\theta_{\bb}(s;h_{0})$ from (\ref{theta bb}) can be rewritten as
\begin{equation}
\label{theta bb non intersecting}
\fl\hspace{15mm}
\boxed{
\theta_{\bb,n}(s;h_{0})=\frac{\P(\mathfrak{b}_{-1}^{s}<h_{0}|\mathfrak{b}_{-n}^{s}<\ldots<\mathfrak{b}_{-1}^{s})}{\P(\mathfrak{b}_{-1}^{s}<\mathfrak{b}_{0}^{s}|\mathfrak{b}_{-n}^{s}<\ldots<\mathfrak{b}_{-1}^{s}\;\text{and}\;\mathfrak{b}_{0}^{s}<\ldots<\mathfrak{b}_{n}^{s})}
}\;.
\end{equation}
Similarly, the coefficient $\tilde{\theta}_{\bb,n}(s;\text{sw})$ from (\ref{theta bb sw}) is equal to 
\begin{equation}
\label{theta bb sw non intersecting}
\fl\hspace{15mm}
\tilde{\theta}_{\bb,n}(s;\text{sw})=\frac{1}{\P(\mathfrak{b}_{-1}^{s}<\mathfrak{b}_{0}^{s}|\mathfrak{b}_{-n}^{s}<\ldots<\mathfrak{b}_{-1}^{s}\;\text{and}\;\mathfrak{b}_{0}^{s}<\ldots<\mathfrak{b}_{n}^{s})}\;.
\end{equation}
which is identical to the denominator in (\ref{theta bb non intersecting}).

Finally, the coefficient $f_{\bb,n}(s;h_{0})$ from (\ref{f bb}) giving the perturbative expansion of the dominant eigenvalue becomes, in this interpretation,
\begin{equation}
\label{f bb non intersecting}
f_{\bb,n}(s;h_{0})=-\langle\mathfrak{b}_{-n}^{s}\,\!\!'(0^{+})\,\mathfrak{b}_{-n}^{s}\,\!\!'(1^{-})\rangle_{\mathfrak{b}_{-n}^{s}<\ldots<\mathfrak{b}_{-1}^{s}<h_{0}}\;,
\end{equation}
where the expectation value is computed with respect to non-intersecting bridges.
\end{subsection}

\begin{subsection}{Conjectures for some expectation values from exact Bethe ansatz formulas}
\label{section conjectures}

At large $L$, the eigenvalue equation for the dominant eigenstate of TASEP gives the cumulant generating function of the height $e(s)$ in terms of Brownian bridges, see (\ref{e[f bb]}) and (\ref{f bb}). Furthermore, the general formula (\ref{theta bb}) for $\theta(s;h_{0})$ is expected to match known explicit formulas for specific initial conditions, that were derived previously, in a non-rigorous way, using singular Euler-Maclaurin asymptotics of the Bethe eigenvectors of TASEP \cite{P2015.2,P2015.3,P2016.1}. This allows us to formulate precise conjectures for some expectation values of Brownian bridges involving extremal values, or equivalently, conditional probabilities of non-intersecting bridges with exponentially distributed distances between the endpoints. Direct probabilistic proofs of these conjectures are still unknown.

\begin{subsubsection}{Dominant eigenvalue.}
\hfill\break
Exact Bethe ansatz calculations \cite{DL1998.1} for TASEP give the dominant eigenvalue as
\begin{equation}
\label{e[chi]}
e(s)=\chi(\nu(s))\;,
\end{equation}
with $\chi$ the polylogarithm
\begin{equation}
\label{chi(v)}
\chi(v)=-\frac{\Li_{5/2}(-\rme^{v})}{\sqrt{2\pi}}\;,
\end{equation}
and $\nu(s)$ the solution of
\begin{equation}
\label{nu(s)}
\chi'(\nu(s))=s\;.
\end{equation}
For $s>0$, (\ref{nu(s)}) defines $\nu(s)$ uniquely since $\chi'$ is a bijection from $\mathbb{R}$ to $\mathbb{R}^{+}$. When $s\to0^{+}$, one has $\nu(s)=\log(\sqrt{2\pi}\,s)+\frac{\sqrt{\pi}}{2}\,s+\mathcal{O}(s^{2})$ and ${e(s)=}\chi(\nu(s))=s+\frac{\sqrt{\pi}}{4}\,s^{2}+\mathcal{O}(s^{3})$. Expanding to the fifth order, we have:
\begin{eqnarray}
&& \frac{s^{2}}{3}+\frac{\chi(\nu(s))}{s}=1+\frac{\sqrt{\pi}}{4}\,s+\Big(\frac{1}{3}+\frac{\pi}{4}-\frac{4\pi}{9\sqrt{3}}\Big)s^{2}\\
&& \hspace{27mm} +\Big(\frac{5}{16}+\frac{3}{8\sqrt{2}}-\frac{1}{\sqrt{3}}\Big)\pi^{3/2}s^{3}+\mathcal{O}(s^{4})\;.\nonumber
\end{eqnarray}
For $s<0$, an analytic continuation is needed \cite{DA1999.1}.

Alternatively, the perturbative expansion of $e(s)$ up to arbitrary order in $s$ can be expressed from (\ref{e[f bb]}) in terms of Brownian bridges (\ref{f bb}). Comparing with (\ref{e[chi]}) leads to
\begin{conjecture}
\label{conjecture master eq}
Let $n$ be a non negative integer and $h_{0}$ an arbitrary regular enough continuous function with $h_{0}(0)=h_{0}(1)=0$. We consider
\begin{equation}
f_{\bb,n}(s;h_{0})=-\frac{\langle b_{n}'(0^{+})\,b_{n}'(1^{-})\,\rme^{-s\max[b_{1}-h_{0}]-s\sum_{j=2}^{n}\max[b_{j}-b_{j-1}]}\rangle_{b_{1},\ldots,b_{n}}}{\langle\rme^{-s\max[b_{1}-h_{0}]-s\sum_{j=2}^{n}\max[b_{j}-b_{j-1}]}\rangle_{b_{1},\ldots,b_{n}}}\;,
\end{equation}
where the averages are taken over independent standard Brownian bridges $b_{1},\ldots,b_{n}$. The perturbative expansion $f_{\bb,n}(s;h_{0})=\frac{s^{2}}{3}+\frac{\chi(\nu(s))}{s}+\mathcal{O}(s^{n})$ is conjectured, with $\chi$ defined in (\ref{chi(v)}) and $\nu(s)$ the solution of (\ref{nu(s)}). This perturbative expansion is in particular independent of $h_{0}$ up to order $s^{n-1}$.
\end{conjecture}

The case $n=0$ is a consequence of $\langle b'(0^{+})\,b'(1^{-})\rangle_{b}=-1$, which follows from the covariance of the Brownian bridge. For $n=2$, using also $\langle\max[b_{1}-b_{0}]\rangle_{b_{0},b_{1}}=\sqrt{\pi}/2$, the conjecture implies $\langle b_{2}'(0^{+})\,b_{2}'(1^{-})\max[b_{2}-b_{1}]\rangle_{b_{1},b_{2}}=-\sqrt{\pi}/4$, which agrees reasonably well with numerical simulations of the Brownian bridge. For $n=3$, additional numerical simulations roughly agree with the conjecture.

\vskip 0.2cm
\noindent
\textbf{Remark:} 
The function $f_{\bb,n}(s;h_{0})$ in conjecture \ref{conjecture master eq} has the alternative expression $f_{\bb,n}(s;h_{0})=-\langle \mathfrak{b}_{-n}^{s}\,\!\!'(0^{+})\,\mathfrak{b}_{-n}^{s}\,\!\!'(1^{-})\rangle_{\mathfrak{b}_{-n}^{s}<\ldots<\mathfrak{b}_{-1}^{s}<h_{0}}$ in terms of the Brownian bridges $\mathfrak{b}_{j}^{s}$ with exponentially distributed distances between the endpoints defined in section \ref{section non intersecting}, and conditioned to never intersect.
\end{subsubsection}

\begin{subsubsection}{Flat initial condition.}
\hfill\break
The flat initial condition corresponds to an initial density profile $\rho_{0}$ of the form (\ref{rho0 fluct}) with $\sigma_{0}(x)=0$, and thus $h_{0}(x)=0$. Comparing with exact results from \cite{P2016.1} suggests that $\theta(s;0)=\theta_{\text{flat}}(s)$ with $\theta_{\text{flat}}$ given by
\begin{eqnarray}
\label{theta flat}
&&\hspace{-12mm} \theta_{\text{flat}}(s)=\frac{s\,\exp(\frac{1}{2}\int_{-\infty}^{\nu(s)}\rmd v\,\chi''(v)^{2})}{(1+\rme^{\nu(s)})^{1/4}\chi''(\nu(s))}\\ 
&& =1+\Big(\frac{\sqrt{\pi}}{2}-\frac{\sqrt{2\pi}}{4}\Big)s+\Big(\frac{1}{4}+\Big(\frac{17}{16}-\frac{1}{2\sqrt{2}}-\frac{4}{3\sqrt{3}}\Big)\pi\Big)s^{2}\\
\label{expan theta flat}
&&\hspace{8mm} +\Big(\Big(\frac{1}{24}-\frac{1}{8\sqrt{2}}\Big)\sqrt{\pi}+\Big(\frac{55}{32}+\frac{9}{32\sqrt{2}}-\frac{4}{\sqrt{3}}+\frac{1}{\sqrt{6}}\Big)\pi^{3/2}\Big)s^{3}
+\mathcal{O}(s^{4})\;.\nonumber 
\end{eqnarray}
The function $\chi$ is defined in (\ref{chi(v)}), and $\nu(s)$ is the solution of (\ref{nu(s)}).

Alternatively, the perturbative expansion of $\theta(s;0)$ up to arbitrary order in $s$ can be expressed from (\ref{theta[theta bb]}) in terms of Brownian bridges (\ref{theta bb}). This leads to the following conjecture:
\begin{conjecture}
\label{conjecture flat}
Let $n$ be a non-negative integer. We consider the function
\begin{equation}
\label{theta flat[bb]}
\hspace{-15mm}
\theta_{\bb,n}(s;0)=\frac{\langle\rme^{-s\sum_{j=1}^{n}\max[b_{j}-b_{j-1}]}\rangle_{b_{0},\ldots,b_{n}}\langle\rme^{-s\max[b_{1}]-s\sum_{j=2}^{n}\max[b_{j}-b_{j-1}]}\rangle_{b_{1},\ldots,b_{n}}}{\langle\rme^{-s\sum_{j=1}^{2n}\max[b_{j}-b_{j-1}]}\rangle_{b_{0},\ldots,b_{2n}}}\;,
\end{equation}
where the averages are taken over independent standard Brownian bridges $b_{j}$. The perturbative expansion $\theta_{\bb,n}(s;0)=\theta_{\text{flat}}(s)+\mathcal{O}(s^{n+1})$ is conjectured, with $\theta_{\text{flat}}(s)$ defined in (\ref{theta flat}).
\end{conjecture}

This conjecture can be checked directly up to order $n=2$ by comparing the expansion (\ref{expan theta flat}) with the expectation values $\langle\max[b]\rangle_{b}=\sqrt{2\pi}/4$, $\langle\max[b]^{2}\rangle_{b}=1/2$, $\langle\max[b_{1}-b_{0}]\rangle_{b_{0},b_{1}}=\sqrt{\pi}/2$, $\langle\max[b_{1}-b_{0}]^{2}\rangle_{b_{0},b_{1}}=1$, $\langle\max[b_{0}]\max[b_{1}-b_{0}]\rangle_{b_{0},b_{1}}=-\frac{1}{2}+\frac{5\pi}{16}$ and $\langle\max[b_{1}-b_{0}]\max[b_{2}-b_{1}]\rangle_{b_{0},b_{1},b_{2}}=-\frac{1}{2}+\frac{2\pi}{3\sqrt{3}}$, which are derived in section \ref{section elementary expectation values max}. For $n=3$, using additional results from section \ref{section elementary expectation values max}, (\ref{theta flat}) leads to $\langle\max[b_{0}]\max[b_{1}-b_{0}]\max[b_{2}-b_{1}]\rangle_{b_{0},b_{1},b_{2}}=-\frac{9\sqrt{\pi}}{16\sqrt{2}}-\frac{41\sqrt{\pi}}{96}+\frac{15\pi^{3/2}}{32\sqrt{2}}\approx0.383683$, which agrees perfectly with the numerics in section \ref{section elementary expectation values max flat}.

\vskip 0.2cm
\noindent
\textbf{Remark:} The function $\theta_{\bb,n}(s;0)$ in conjecture \ref{conjecture flat} has the alternative expression $\theta_{\bb,n}(s;0)=\frac{\P(\mathfrak{b}_{-1}^{s}<0|\mathfrak{b}_{-n}^{s}<\ldots<\mathfrak{b}_{-1}^{s})}{\P(\mathfrak{b}_{-1}^{s}<\mathfrak{b}_{0}^{s}|\mathfrak{b}_{-n}^{s}<\ldots<\mathfrak{b}_{-1}^{s}\;\text{and}\;\mathfrak{b}_{0}^{s}<\ldots<\mathfrak{b}_{n}^{s})}$ in terms of the non-intersecting Brownian bridges $\mathfrak{b}_{j}^{s}$ defined in section \ref{section non intersecting}.
\end{subsubsection}

\begin{subsubsection}{Stationary initial condition.}
\hfill\break
The stationary initial condition corresponds to an initial density profile $\rho_{0}$ of the form (\ref{rho0 fluct}), but with a corresponding $h_{0}$ random and equal in law to a standard Brownian bridge $b$. The coefficient $\theta$ in the moment generating function (\ref{GFh[theta] fluct}) of $h(x,t)$ for stationary initial condition is then equal to the average $\langle\theta(s;b)\rangle_{b}$. Exact results from \cite{P2016.1} suggest that $\langle\theta(s;b)\rangle_{b}=\theta_{\text{stat}}(s)$ with $b$ a standard Brownian bridge and $\theta_{\text{stat}}$ given by
\begin{eqnarray}
\label{theta stat}
\theta_{\text{stat}}(s)&=& \frac{\sqrt{2\pi}\,s^{2}\exp(\int_{-\infty}^{\nu(s)}\rmd v\,\chi''(v)^{2})}{\rme^{\nu(s)}\chi''(\nu(s))} \\
 &=& 1+\Big(\frac{1}{2}+\Big(\frac{1}{4}-\frac{2}{3\sqrt{3}}\Big)\pi\Big)s^{2}\\
&& \,\,\, +\Big(-\frac{\sqrt{\pi}}{6}+\Big(\frac{1}{2}+\frac{1}{\sqrt{2}}-\frac{2}{\sqrt{3}}\Big)\pi^{3/2}\Big)s^{3}+\mathcal{O}(s^{4})\;.\nonumber
\end{eqnarray}
The function $\chi$ is defined in (\ref{chi(v)}), and $\nu(s)$ is the solution of (\ref{nu(s)}).

Alternatively, the perturbative expansion of $\langle\theta(s;b)\rangle_{b}$ up to arbitrary order in $s$ can be expressed from (\ref{theta[theta bb]}) in terms of Brownian bridges (\ref{theta bb}). This leads to the following conjecture:
\begin{conjecture}
\label{conjecture stat}
Let $n$ be a non-negative integer. We consider the function
\begin{eqnarray}
\label{theta stat[bb]}
&&\langle\theta_{\bb,n}(s;b)\rangle_{b}=\frac{\big(\langle\rme^{-s\sum_{j=1}^{n}\max[b_{j}-b_{j-1}]}\rangle_{b_{0},\ldots,b_{n}}\big)^{2}}{\langle\rme^{-s\sum_{j=1}^{2n}\max[b_{j}-b_{j-1}]}\rangle_{b_{0},\ldots,b_{2n}}}\;,
\end{eqnarray}
where the averages are taken over independent standard Brownian bridges $b_{j}$. The perturbative expansion $\langle\theta_{\bb,n}(s;b)\rangle_{b}=\theta_{\text{stat}}(s)+\mathcal{O}(s^{n+1})$ is conjectured, with $\theta_{\text{stat}}(s)$ defined in (\ref{theta stat}).
\end{conjecture}

This conjecture can be checked directly up to $n=2$ by using the expectation values $\langle\max[b_{1}-b_{0}]\rangle_{b_{0},b_{1}}=\sqrt{\pi}/2$ and $\langle\max[b_{1}-b_{0}]\max[b_{2}-b_{1}]\rangle_{b_{0},b_{1},b_{2}}=-\frac{1}{2}+\frac{2\pi}{3\sqrt{3}}$ derived in section \ref{section elementary expectation values max stat}. For $n=3$, using additional results from section \ref{section elementary expectation values max stat}, (\ref{theta stat}) leads to $\langle\max[b_{1}-b_{0}]\max[b_{2}-b_{1}]\max[b_{3}-b_{2}]\rangle_{b_{0},b_{1},b_{2},b_{3}}=-\frac{19\sqrt{\pi}}{24}+\frac{\pi^{3/2}}{2\sqrt{2}}\approx0.565509$, which agrees perfectly with the numerics in section \ref{section elementary expectation values max stat}.

\vskip 0.2cm
\noindent
\textbf{Remark:} The function $\langle\theta_{\bb,n}(s;b)\rangle_{b}$ in conjecture \ref{conjecture stat} has the alternative expression $\langle\theta_{\bb,n}(s;b)\rangle_{b}=\frac{\P(\mathfrak{b}_{-1}^{s}<\mathfrak{b}_{0}^{s}|\mathfrak{b}_{-n}^{s}<\ldots<\mathfrak{b}_{-1}^{s})}{\P(\mathfrak{b}_{-1}^{s}<\mathfrak{b}_{0}^{s}|\mathfrak{b}_{-n}^{s}<\ldots<\mathfrak{b}_{-1}^{s}\;\text{and}\;\mathfrak{b}_{0}^{s}<\ldots<\mathfrak{b}_{n}^{s})}$ in terms of the non-intersecting Brownian bridges $\mathfrak{b}_{j}^{s}$ defined in section \ref{section non intersecting}.
\end{subsubsection}

\begin{subsubsection}{Domain wall initial condition.}
\label{section dw}
\hfill\break
Domain wall initial condition corresponds to the finite density profile $\rho_{0}(x)=\openone_{\{0\leq x\leq\rhobar\}}$ and to the height profile $\mathcal{H}_{0}$ from (\ref{H0[rho0]}), $\mathcal{H}_{0}(x)=-(1-\rhobar)x\openone_{\{0\leq x\leq\rhobar\}}+\rhobar(x-1)\openone_{\{\rhobar<x\leq1\}}$ for $x$ in the interval $[0,1]$. The global minimum $\min[\mathcal{H}_{0}]=-\rhobar(1-\rhobar)$ is reached only once, at $x=\rhobar$. Using (\ref{h(x,t;H/epsilon) -> htilde(x,t;H)}) and (\ref{theta(s;H/epsilon) -> thetatilde(s;H)}), exact results in \cite{P2016.1} for domain wall initial condition suggest $\tilde{\theta}(s;\text{sw})=\tilde{\theta}_{\text{dw}}(s)$ with $\tilde{\theta}_{\text{dw}}$ given by
\begin{eqnarray}
\label{theta dw}
\tilde{\theta}_{\text{dw}}(s)&=&\frac{s\,\exp(\int_{-\infty}^{\nu(s)}\rmd v\,\chi''(v)^{2})}{\chi''(\nu(s))}\\
&=& 1+\frac{\sqrt{\pi}}{2}\,s
+\Big(\frac{1}{2}+\Big(\frac{3}{4}-\frac{4}{3\sqrt{3}}\Big)\pi\Big)s^{2}\\
&&\hspace{17mm} +\Big(\frac{\sqrt{\pi}}{12}+\Big(\frac{5}{4}+\frac{3}{2\sqrt{2}}-\frac{4}{\sqrt{3}}\Big)\pi^{3/2}\Big)s^{3}+\mathcal{O}(s^{4})\;.\nonumber
\end{eqnarray}
The function $\chi$ is defined in (\ref{chi(v)}), and $\nu(s)$ is the solution of (\ref{nu(s)}).

Alternatively, the perturbative expansion of $\tilde{\theta}(s;\text{sw})$ up to arbitrary order in $s$ can be expressed from (\ref{theta sw[theta bb sw]}) in terms of Brownian bridges (\ref{theta bb sw}). This leads to the following conjecture:
\begin{conjecture}
\label{conjecture dw}
Let $n$ be a non-negative integer. We consider the function
\begin{equation}
\label{theta dw[bb]}
\tilde{\theta}_{\bb,n}(s;\text{sw})=\frac{\langle\rme^{-s\sum_{j=1}^{n}\max[b_{j}-b_{j-1}]}\rangle_{b_{0},\ldots,b_{n}}\langle\rme^{-s\sum_{j=2}^{n}\max[b_{j}-b_{j-1}]}\rangle_{b_{1},\ldots,b_{n}}}{\langle\rme^{-s\sum_{j=1}^{2n}\max[b_{j}-b_{j-1}]}\rangle_{b_{0},\ldots,b_{2n}}}\;,
\end{equation}
where the averages are taken over independent standard Brownian bridges $b_{j}$. The perturbative expansion $\tilde{\theta}_{\bb,n}(s;\text{sw})=\tilde{\theta}_{\text{dw}}(s)+\mathcal{O}(s^{n+1})$ is conjectured, with $\tilde{\theta}_{\text{dw}}(s)$ defined in (\ref{theta dw}).
\end{conjecture}

This conjecture can be checked directly up to $n=2$ by using the expectation values $\langle\max[b_{1}-b_{0}]\rangle_{b_{0},b_{1}}=\sqrt{\pi}/2$, $\langle\max[b_{1}-b_{0}]^{2}\rangle_{b_{0},b_{1}}=1$ and $\langle\max[b_{1}-b_{0}]\max[b_{2}-b_{1}]\rangle_{b_{0},b_{1},b_{2}}=-\frac{1}{2}+\frac{2\pi}{3\sqrt{3}}$ derived in section \ref{section elementary expectation values max}. For $n=3$, using additional results from section \ref{section elementary expectation values max stat}, (\ref{theta dw}) leads again to $\langle\max[b_{1}-b_{0}]\max[b_{2}-b_{1}]\max[b_{3}-b_{2}]\rangle_{b_{0},b_{1},b_{2},b_{3}}=-\frac{19\sqrt{\pi}}{24}+\frac{\pi^{3/2}}{2\sqrt{2}}\approx0.565509$, which agrees perfectly with the numerics in section \ref{section elementary expectation values max stat}.

\vskip 0.2cm
\noindent
\textbf{Remark:} The function $\tilde{\theta}_{\bb,n}(s;\text{sw})$ in conjecture \ref{conjecture dw} has the alternative expression $\tilde{\theta}_{\bb,n}(s;\text{sw})=1/\P(\mathfrak{b}_{-1}^{s}<\mathfrak{b}_{0}^{s}|\mathfrak{b}_{-n}^{s}<\ldots<\mathfrak{b}_{-1}^{s}\;\text{and}\;\mathfrak{b}_{0}^{s}<\ldots<\mathfrak{b}_{n}^{s})$ in terms of the non-intersecting Brownian bridges $\mathfrak{b}_{j}^{s}$ defined in section \ref{section non intersecting}.
\end{subsubsection}

\begin{subsubsection}{Ratio between stationary and domain wall initial condition.}
\hfill\break
We observe that the ratio of (\ref{theta stat}) and (\ref{theta dw}) has a simple expression involving only $\nu(s)$. From conjectures \ref{conjecture stat} and \ref{conjecture dw}, this ratio can be expressed in terms of Brownian bridges as
\begin{equation}
\frac{\langle\rme^{-s\sum_{j=1}^{n}\max[b_{j}-b_{j-1}]}\rangle_{b_{0},\ldots,b_{n}}}{\langle\rme^{-s\sum_{j=2}^{n}\max[b_{j}-b_{j-1}]}\rangle_{b_{1},\ldots,b_{n}}}=\sqrt{2\pi}\,s\,\rme^{-\nu(s)}+\mathcal{O}(s^{n+1})\;,
\end{equation}
with
\begin{equation}
\fl\hspace{5mm}
\sqrt{2\pi}\,s\,\rme^{-\nu(s)}=1-\frac{\sqrt{\pi}}{2}\,s+\Big(\frac{2\pi}{3\sqrt{3}}-\frac{\pi}{4}\Big)s^{2}+\Big(\frac{1}{\sqrt{3}}-\frac{1}{4}-\frac{1}{2\sqrt{2}}\Big)\pi^{3/2}s^{3}+\mathcal{O}(s^{4})\;.
\end{equation}
Alternatively, using the non-intersecting Brownian bridges $\mathfrak{b}_{j}^{s}$ defined in section \ref{section non intersecting}, one has
\begin{equation}
\P(\mathfrak{b}_{-1}^{s}<\mathfrak{b}_{0}^{s}|\mathfrak{b}_{-n}^{s}<\ldots<\mathfrak{b}_{-1}^{s})=\sqrt{2\pi}\,s\,\rme^{-\nu(s)}+\mathcal{O}(s^{n+1})\;.
\end{equation}
\end{subsubsection}

\end{subsection}

\begin{subsection}{Conclusions}
We have derived in this article a relation between late time KPZ fluctuations in finite volume with periodic boundaries and non-intersecting Brownian bridges with exponentially distributed distances between the endpoints is obtained. This relation is obtained, after taking a continuous limit, from the exact matrix product representation \cite{LM2011.1,L2013.1} for the dominant eigenstate of a deformed Markov operator counting the current in TASEP. It would be desirable to find a direct probabilistic derivation in the continuum, starting already with the KPZ equation (or its Cole-Hopf transform, the stochastic heat equation with multiplicative noise).

The relation between KPZ fluctuations and non-intersecting Brownians studied in this paper is reminiscent of the interpretation of the Airy$_{2}$ process, describing spatial correlations at a given time of an infinitely long KPZ interface with nonzero curvature, as a large $n$ scaling limit of Dyson's $n\times n$ matrix Brownian motion \cite{D1962.1}, whose $n$ eigenvalues are equal in law to Wiener processes conditioned to never intersect \cite{G1999.1}. The study of non-intersecting Wiener processes in various configurations has been very active in the past few years \cite{N-ISC}, especially in relation with KPZ universality.

Using recent asymptotic calculations \cite{P2016.1} of Bethe eigenstates for TASEP, the connection to Brownian bridges studied in this paper additionally provides nice exact formulas (\ref{theta flat}), (\ref{theta stat}), (\ref{theta dw}) involving polylogarithms for a few conditional probabilities related to non-intersecting Brownian bridges, or equivalently, for expectation values involving maxima of Brownian bridges. Precise statements are formulated in section \ref{section conjectures}. Again, direct proofs of these formulas would be welcome, as well as extensions to more general initial states, especially perturbative expansions for initial heights with either large or small amplitude.

Finally, generalization of the Brownian bridge formulas to KPZ fluctuations in an interval with open boundaries, where the stationary state is no longer a Brownian bridge but instead the sum of a Brownian bridge plus an independent Brownian excursion \cite{DEL2004.1}, are definitely worth investigating. Besides, extensions to the higher excited states appearing in the dynamical partition function (\ref{GFh[theta_r] fluct}) would shed light on the relaxation process.\\

\noindent
{\bf Acknowledgement:} This work was granted access to the HPC resources of CALMIP supercomputing center under the allocation 2018-P18003.
\end{subsection}
\end{section}

\begin{section}{From the matrix product representation to Brownian bridges}
\label{section matrix product -> bb}

In this section, we explain how the matrix product representation of \cite{LM2011.1,L2013.1} can be reformulated in terms of height functions, and show that large $L$ asymptotics lead to the Brownian bridge formulas of section \ref{section theta bb}.

\begin{subsection}{Deformed Markov operator and height fluctuations}
Let $\Omega$ be the set of configurations (micro-states) of the periodic TASEP with $L$ sites and $N$ particles, of cardinal $|\Omega|=\C{L}{N}$, and $P_{\tm}(\mathcal{C})$ the probabilities of the configurations $\mathcal{C}\in\Omega$ at time $\tm$. Since TASEP is a Markov (i.e. memoryless) process, the evolution in time of the probability vector $|P(\tm)\rangle=\sum_{\mathcal{C}\in\Omega}P_{\tm}(\mathcal{C})|\mathcal{C}\rangle$ is given by the master equation
\begin{equation}
\frac{\rmd}{\rmd\tm}|P(\tm)\rangle=M|P(\tm)\rangle\;,
\end{equation}
where $M$ is the Markov operator.

The configurations $\mathcal{C}\in\Omega$ do not keep track of the number of particles that have hopped from a given site $i$ to the next site $i+1$ up to time $\tm$. In order to characterize the fluctuations of the height function $H_{i}(\tm)$, a local deformation $M_{i}(\gamma)$ of the Markov operator is needed \cite{DL1998.1}. This deformation is built by multiplying the elements of $M$ corresponding to transitions from the site $i$ to the site $i+1$ by the factor $\rme^{\gamma}$, where the deformation parameter $\gamma$ is a fugacity conjugate to the height.

The generating function of the height $\langle\rme^{\gamma H_{i}(\tm)}\rangle$ can be expanded over the eigenstates of $M_{i}(\gamma)$, as recalled in \ref{appendix sum eigenstates}. The universal KPZ statistics at large $L$ on the time scale (\ref{tKPZ}) are then formulated in terms of the rescaled fugacity $s$ defined by
\begin{equation}
\gamma=\frac{s}{\sqrt{\rhobar(1-\rhobar)L}}\;.
\end{equation}
On this time scale, only the eigenstates of $M_{i}(\gamma)$ with eigenvalue $E_{r}(\gamma)$ (which is independent of $i$, see \ref{appendix sum eigenstates}) such that $\Re(E_{r}(\gamma)-E_{0}(\gamma))\sim L^{-3/2}$ contribute. We denote by $|\Psi_{r}^{0}(\gamma)\rangle$ and $\langle\Psi_{r}^{0}(\gamma)|$ the corresponding right and left eigenvectors.

The functions $e_{r}(s)$ from (\ref{GFh[theta_r] fluct}), (\ref{GFh[theta_r] finite}) correspond to the term of order $L^{-3/2}$ in the large $L$ asymptotics of the eigenvalues $E_{r}(\gamma)$, as shown in (\ref{GFH[psi0]}): 
\begin{equation}
\label{E_r[chi_r]}
\frac{E_{r}(\gamma)-\rhobar(1-\rhobar)\gamma}{\sqrt{\rhobar(1-\rhobar)}}\simeq-\frac{\rmi(1-2\rhobar)p_{r}}{L}+\frac{e_{r}(s)}{L^{3/2}}\;.
\end{equation}
The dominant eigenvalue $e(s)=e_{0}(s)$ is given by (\ref{e[chi]}) and explicit formulas for all the coefficients $e_{r}(s)$ are also known \cite{P2014.1}.

Furthermore, for an initial condition of the form (\ref{rho0 fluct}), and the corresponding height profile $h_{0}$ defined in (\ref{h0[sigma0]}), the coefficients $\theta_{r}(s;h_{0})$ in (\ref{GFh[theta_r] fluct}) are obtained using (\ref{H[h] fluct}), (\ref{E_r[chi_r]}) and (\ref{GFH[psi0]}) as
\begin{equation}
\label{theta_r[psi_r] fluct}
\theta_{r}(s;h_{0})=\lim_{L\to\infty}\frac{(\sum_{\mathcal{C}\in\Omega}\langle\mathcal{C}|\Psi_{r}^{0}(\gamma)\rangle)\,\langle\Psi_{r}^{0}(\gamma)|P_{0}\rangle}{\langle\Psi_{r}^{0}(\gamma)|\Psi_{r}^{0}(\gamma)\rangle}\;,
\end{equation}
with $P_{0}$ the initial state. Exact results for flat initial conditions \cite{P2015.3} (where $X_{j}(0)=j/\rhobar$, with $\rhobar^{-1}$ integer) and stationary initial conditions\cite{P2016.1} (where all $\mathcal{C}$ in $\Omega$ have the same weight $1/|\Omega|$) as well as high precision Bethe ansatz numerics for a few other initial states, confirm that the large $L$ limit in (\ref{theta_r[psi_r] fluct}) is well defined.

Similarly, for a finite initial density profile $\rho_{0}$, with corresponding height profile $\mathcal{H}_{0}$ defined in (\ref{H0[rho0]}), the coefficients $\tilde{\theta}_{r}(s;\mathcal{H}_{0})$ in (\ref{GFh[theta_r] finite}) are given by, using (\ref{H[h] finite}), (\ref{E_r[chi_r]}) and (\ref{GFH[psi0]}), 
\begin{equation}
\label{theta_r[psi_r] finite}
\tilde{\theta}_{r}(s;\mathcal{H}_{0})=\lim_{L\to\infty}\rme^{-L^{2}\gamma\min[\mathcal{H}_{0}]}\frac{\Big(\sum_{\mathcal{C}\in\Omega}\langle\mathcal{C}|\Psi_{r}^{0}(\gamma)\rangle\Big)\langle\Psi_{r}^{0}(\gamma)|P_{0}\rangle}{\langle\Psi_{r}^{0}(\gamma)|\Psi_{r}^{0}(\gamma)\rangle}\;.
\end{equation}
Exact Bethe ansatz results for domain wall \cite{P2015.2} initial condition $X_{j}(0)=j$, $j=1,\ldots,N$, high precision extrapolation in some cases with piecewise constant density profile with more domain walls, and additional numerics for a few other cases confirm again that the large $L$ limit in (\ref{theta_r[psi_r] finite}) is well defined.
\end{subsection}

\begin{subsection}{Dominant eigenvector: matrix product and height representations}
\label{section matrix product}
The left and right dominant eigenvectors of $M_{i}(\gamma)$ have a matrix product representation found in \cite{LM2011.1,L2013.1}. The alternative formulation introduced below in terms of height functions is the key to the asymptotic analysis performed in this paper, which leads to Brownian bridges.

\begin{subsubsection}{Matrix product representation.}
\hfill\break
The dominant eigenvalue of the (non-deformed) Markov operator $M=M_{0}(0)$ is equal to zero. The corresponding left and right dominant eigenvectors are represented in configuration basis by
\begin{equation}
\label{0L 0R}
\langle0|=\frac{1}{\Omega}\sum_{\mathcal{C}}\langle\mathcal{C}|
\qquad\text{and}\qquad
|0\rangle=\frac{1}{\Omega}\sum_{\mathcal{C}}|\mathcal{C}\rangle\;.
\end{equation}
The first equation follows directly from the Markov property, while the second one is a consequence of pairwise balance \cite{D1998.1}.

The dominant eigenvalue of the deformed operator $M_{0}(\gamma)$ is no longer zero when $\gamma\neq0$. The dominant eigenvalue and the corresponding eigenvectors of $M_{0}(\gamma)$ were obtained in \cite{LM2011.1,L2013.1} (though these papers deal mainly with the more complicated case of the asymmetric exclusion process with open boundaries, the periodic case is mentioned in \cite{L2013.1}).

The dominant eigenvectors $\langle\Psi_{0}^{0}(\gamma)|$ and $|\Psi_{0}^{0}(\gamma)\rangle$ of $M_{0}(\gamma)$ are constructed perturbatively, with respect to $\gamma$, by repeated action on the vectors given in (\ref{0L 0R}) of a transfer operator $T(\gamma)$ which commutes with $M_{0}(\gamma)$ and such that $T(0)\propto|0\rangle\langle0|$. For any non-negative integer $n$, one has
\begin{eqnarray}
\label{psiL[T]}
&& \langle\Psi_{0}^{0}(\gamma)|=\langle0|T(\gamma)^{n}+\mathcal{O}(\gamma^{n+1})\\
\label{psiR[T]}
&& |\Psi_{0}^{0}(\gamma)\rangle=T(\gamma)^{n}|0\rangle+\mathcal{O}(\gamma^{n+1})\;.
\end{eqnarray}
In the canonical basis, the transfer operator has the following matrix product representation
\begin{equation}
\label{T[ADE]}
\langle\mathcal{C}'|T(\gamma)|\mathcal{C}\rangle=\tr[AX_{n_{1}',n_{1}}X_{n_{2}',n_{2}}\ldots X_{n_{L}',n_{L}}]\;,
\end{equation}
where the configurations $\mathcal{C}$ and $\mathcal{C}'$ are represented by their respective occupation numbers $n_{i}$ and $n'_{i}$. The operators $X_{n',n}$ are
\begin{equation}
\label{X[1DE]}
X_{0,0}=X_{1,1}=\openone\;,
\quad
X_{1,0}=D\;,
\quad
X_{0,1}=E\;,
\end{equation}
where $A$, $D$ and $E$ verify the algebra
\begin{equation}
\label{algebra ADE}
DA=\rme^{-\gamma}AD\;,
\quad
AE=\rme^{-\gamma}EA\;,
\quad
DE=\openone\;,
\end{equation}
and the normalization $\tr[A]=1$ is chosen.

A similar transfer operator structure was also used in \cite{PEM2009.1} to generate the dominant eigenstate of the (non-deformed) Markov operator of an exclusion process involving more species of particles. There, $\rme^{\gamma}$ was replaced by the ratio between forward and backward hopping rates.
\end{subsubsection}

\begin{subsubsection}{Height representation of the operator $T(\gamma)$.}
\hfill\break
The starting point of all the calculations in the present paper is the alternative representation
\begin{equation}
\label{T[H]}
\langle\mathcal{C}'|T(\gamma)|\mathcal{C}\rangle=\exp\Big(-\gamma\max_{1\leq i\leq L}(H_{i}'-H_{i})\Big)
\end{equation}
for the transfer operator in terms of the height function $H_{i}=\sum_{k=1}^{i}(\rhobar-n_{k})$, $H_{0}=H_{L}=0$, and similarly for $H_{i}'$ in terms of $n_{k}'$.

The expression (\ref{T[H]}) can be derived graphically from (\ref{T[ADE]}) by plotting $H_{i}'-H_{i}=\sum_{k=1}^{i}(n_{k}-n_{k}')$ as a function of the site $i=1,\ldots,L$. After connecting the dots, each section of the graph between neighbouring sites is associated to an operator in the matrix product representation (\ref{T[ADE]}), (\ref{X[1DE]}): the horizontal sections of the graph correspond to $\openone$, the increasing sections to $E$ and the decreasing sections to $D$.
\begin{center}
\setlength{\unitlength}{0.8mm}
\begin{picture}(145,42)
  \setcounter{X}{0}\setcounter{Y}{20}
  \put(0,\theY){\color[rgb]{0.7,0.7,0.7}\vector(1,0){145}}
  \put(144,21.5){$i$}
  \put(-1,15){$0$}\put(9,15){$1$}\put(19,15){$2$}\put(28,15){$\ldots$}\put(138,15){$L$}
  \up\hz\up\down\down\hz\down\down\hz\up\up\hz\up\down
\end{picture}
\end{center}
Since $H_{0}=H_{L}=H_{0}'=H_{L}'=0$, the number of operators $D$ and $E$ is the same. Removing horizontal sections of the graph does not change the value of the matrix product $\tr[AX_{n_{1}',n_{1}}X_{n_{2}',n_{2}}\ldots X_{n_{L}',n_{L}}]$. Furthermore, the algebra $DE=\openone$ ensures that one can also erase a decreasing section followed by an increasing section of the same length without changing the matrix product. For the above example, this procedure leads to
\begin{center}
\setlength{\unitlength}{0.8mm}
\begin{picture}(40,22)
  \setcounter{X}{0}\setcounter{Y}{0}
  \put(0,\theY){\color[rgb]{0.7,0.7,0.7}\line(1,0){40}}
  \up\up\down\down
\end{picture}
\end{center}
We observe that the erasing procedure does not change the maximum $m=\max_{1\leq i\leq L}(H_{i}'-H_{i})$ of the curve, and always leads to the matrix product $\langle\mathcal{C}'|T(\gamma)|\mathcal{C}\rangle=\tr[AE^{m}D^{m}]$. The algebra (\ref{algebra ADE}) together with the normalization $\tr[A]=1$ finally leads to (\ref{T[H]}).
\end{subsubsection}

\end{subsection}

\begin{subsection}{Coefficients $\theta$ and $\tilde{\theta}$ of the height generating functions at late time}
Combining the expressions (\ref{theta_r[psi_r] fluct}) and (\ref{theta_r[psi_r] finite}) for $\theta(s;h_{0})$ and $\tilde{\theta}(s;\mathcal{H}_{0})$ with the matrix product representation (\ref{psiL[T]}), (\ref{psiR[T]}) of the dominant eigenvector, we obtain a perturbative expansion up to arbitrary order $n$ in $s$ of the coefficients $\theta$ and $\tilde{\theta}$:
\begin{eqnarray}
\label{theta[T] fluct}
&&
\fl\hspace{15mm}
\theta(s;h_{0})=\lim_{L\to\infty}\frac{\sum_{\mathcal{C}\in\Omega}\langle\mathcal{C}|T(\gamma)^{n}|0\rangle\langle0|T(\gamma)^{n}|P_{0}\rangle}{\langle0|T(\gamma)^{2n}|0\rangle}+\mathcal{O}(s^{n+1})\\
\label{theta[T] finite}
&&
\fl\hspace{15mm}
\tilde{\theta}(s;\mathcal{H}_{0})=\lim_{L\to\infty}\rme^{-L^{2}\gamma\min[\mathcal{H}_{0}]}\frac{\sum_{\mathcal{C}\in\Omega}\langle\mathcal{C}|T(\gamma)^{n}|0\rangle\langle0|T(\gamma)^{n}|P_{0}\rangle}{\langle0|T(\gamma)^{2n}|0\rangle}+\mathcal{O}(s^{n+1})\;.
\end{eqnarray}

The large $L$ limit in (\ref{theta[T] fluct}) and (\ref{theta[T] finite}) can be performed by first inserting the decomposition of the identity $\openone=\sum_{\mathcal{C}\in\Omega}|\mathcal{C}\rangle\langle\mathcal{C}|$ between the operators $T(\gamma)$ and interpreting the configurations $\mathcal{C}$ in height representation as random walks. In the scaling limit, these random walks then converge to Brownian bridges by Donsker's theorem (see e.g. \cite{D1999.1}; for completeness sake, a derivation of this fact is provided in \ref{appendix TASEP -> bb}). Combining (\ref{theta[T] fluct}), (\ref{box <f>[b] leading order}) and (\ref{T[b]}) yields (\ref{theta[theta bb]}) and (\ref{theta bb}). Similarly, combining (\ref{theta[T] finite}), (\ref{box <f>[b] leading order}) and (\ref{T[b]}) leads to (\ref{theta sw[theta bb sw]}) and (\ref{theta bb sw}).
\end{subsection}

\begin{subsection}{The Derrida-Lebowitz large deviation function $e(s)$}
The eigenvalue equation for the left dominant eigenvector of $M_{0}(\gamma)$ reads $\langle\Psi_{0}(\gamma)|M_{0}(\gamma)|P_{0}\rangle=E_{0}(\gamma)\langle\Psi_{0}(\gamma)|P_{0}\rangle$, with $|P_{0}\rangle$ an arbitrary vector. Using (\ref{psiL[T]}), (\ref{0L 0R}) and the fact that $M_{0}(\gamma)$ and $T(\gamma)$ commute, one has for any non-negative integer $n$
\begin{equation}
\label{eigenvalue eq tmp}
\sum_{\mathcal{C}'\in\Omega}\langle\mathcal{C}'|M_{0}(\gamma)T(\gamma)^{n}|P_{0}\rangle=E_{0}(\gamma)\sum_{\mathcal{C}\in\Omega}\langle\mathcal{C}|T(\gamma)^{n}|P_{0}\rangle+\mathcal{O}(\gamma^{n+1})\;.
\end{equation}
Next, we remark that the deformed operator $M_{0}(\gamma)$ differs from $M=M_{0}(0)$ only through boundary terms. Besides, since $M$ is a Markov operator, $\sum_{\mathcal{C}}\langle\mathcal{C}|$ is the left eigenvector of $M$ with eigenvalue $0$. This leads to the relation
\begin{equation}
\label{<1|M0(gamma)}
\sum_{\mathcal{C}'\in\Omega}\langle\mathcal{C}'|M_{0}(\gamma)|\mathcal{C}\rangle=(\rme^{\gamma}-1)1_{\{n_{1}=0\}}1_{\{n_{L}=1\}}\;,
\end{equation}
where the $n_{i}$'s are the occupation numbers for the configuration $\mathcal{C}$.
Inserting the identity $\openone=\sum_{\mathcal{C}\in\Omega}|\mathcal{C}\rangle\langle\mathcal{C}|$ between $M_{0}(\gamma)$ and $T(\gamma)^{n}$ in the left side of (\ref{eigenvalue eq tmp}), and using (\ref{<1|M0(gamma)}) in order to eliminate the summation over $\mathcal{C}'$, we finally obtain
\begin{equation}
\label{eigenvalue eq[T]}
\fl\hspace{15mm}
\sum_{\mathcal{C}\in\Omega}1_{\{n_{1}=0\}}1_{\{n_{L}=1\}}\langle\mathcal{C}|T(\gamma)^{n}|P_{0}\rangle=\frac{E_{0}(\gamma)}{\rme^{\gamma}-1}\,\sum_{\mathcal{C}\in\Omega}\langle\mathcal{C}|T(\gamma)^{n}|P_{0}\rangle+\mathcal{O}(\gamma^{n+1})\;.
\end{equation}
In the left hand side, the $n_{i}$'s are the occupation numbers of the configuration $\mathcal{C}$. The expression (\ref{eigenvalue eq[T]}) does not involve $M_{0}(\gamma)$ any more, which makes it particularly suitable for large $L$ asymptotic analysis. The derivation of (\ref{e[f bb]}), (\ref{f bb}) from (\ref{eigenvalue eq[T]}) is worked out in \ref{section bb master equation} using a sub-leading correction to Donsker's theorem.
\end{subsection}

\end{section}

\begin{section}{Calculation of some Brownian averages involving extremal values}
\label{section expectation values bb}

In this section, we compute some Brownian expectation values that are relevant to obtain the first cumulants of the KPZ height. We perform perturbative expansions for initial conditions with small and large amplitudes. We also obtain exact results in the case the initial height fonction is piecewise-linear or is parabolic.

\begin{subsection}{Elementary expectation values}
\label{section elementary expectation values max}

\begin{subsubsection}{Moments of $\max[b]$.}
\label{section GF max b}
\hfill\break
The probability distribution of the maximum of a Brownian bridge $b$ can be computed by the method of images. One has
\begin{equation}
\P(\max[b]<z)=(1-\rme^{-2z^{2}})\openone_{\{z>0\}}\;,
\end{equation}
which implies the classical result
\begin{equation}
\label{<max[b]^r>}
\langle\max[b]^{r}\rangle_{b}=2^{-r/2}\,\Gamma(1+r/2)\;.
\end{equation}
The corresponding generating function involves the error function:
\begin{equation}
\langle\rme^{-s\max[b]}\rangle_{b}=1-\frac{\sqrt{2\pi}}{4}\,s\,\rme^{s^{2}/8}\Big(1-\erf\Big(\frac{s}{2\sqrt{2}}\Big)\Big)\;.
\end{equation}
\end{subsubsection}

\begin{subsubsection}{Brownian averages for stationary and domain wall initial condition.}
\label{section elementary expectation values max stat}
\hfill\break
We compute in this section a few Brownian averages related to (\ref{theta stat[bb]}) and (\ref{theta dw[bb]}), which provide a partial check of conjectures \ref{conjecture stat} and \ref{conjecture dw}.

The expectation value $\langle\max[b_{1}-b_{0}]^{r}\rangle_{b_{0},b_{1}}$ reduces to $\langle\max[b]\rangle_{b}$ since $(b_{1}-b_{0})/\sqrt{2}$ is equal in law to a standard Brownian bridge. Thus,
\begin{equation}
\label{<max[b1-b0]^r>}
\langle\max[b_{1}-b_{0}]^{r}\rangle_{b_{0},b_{1}}=\Gamma(1+r/2)
\end{equation}
and
\begin{equation}
\langle\rme^{-s\max[b_{1}-b_{0}]}\rangle_{b_{0}}=1+\frac{\sqrt{\pi}}{2}\,s\,\rme^{s^{2}/4}\Big(-1+\erf\Big(\frac{s}{2}\Big)\Big)\;.
\end{equation}

For more general correlation functions of the form $\langle\max[b_{1}-b_{0}]^{r_{1}}\max[b_{2}-b_{1}]^{r_{2}}\ldots\rangle_{b_{0},b_{1},b_{2},\ldots}$ with independent standard Brownian bridges $b_{j}$, we use
\begin{eqnarray}
&&
\fl\hspace{5mm}
\P(b_{1}-b_{0}<z_{1},\ldots,b_{n}-b_{n-1}<z_{n})\\
&&
\fl\hspace{5mm}
=\frac{\P(w_{1}-w_{0}<z_{1},\ldots,w_{n}-w_{n-1}<z_{n},0<w_{0}(1)<\rmd y_{0},\ldots,0<w_{n}(1)<\rmd y_{n})}{\P(0<w_{0}(1)<\rmd y_{0},\ldots,0<w_{n}(1)<\rmd y_{n})}\;,\nonumber
\end{eqnarray}
where $w_{j}$, $j=0,\ldots,n$ are independent Wiener processes and all $\rmd y_{j}\to0$. The denominator is equal to $(\sqrt{2\pi})^{-n-1}\rmd y_{0}\ldots\rmd y_{n}$. The numerator is the distribution of a process of $n+1$ Brownian motions that terminates when two particles collide, for which the Karlin-McGregor formula \cite{KMG1959.1} gives a determinant. Alternatively, the numerator can be thought of as a $n+1$ dimensional Brownian motion in the Weyl chamber associated to root system $A_{n}$ with absorbing boundaries, and equation (6) of \cite{G1999.1} gives
\begin{equation}
\label{P bn stat}
\fl\hspace{20mm}
\P(b_{1}-b_{0}<z_{1},\ldots,b_{n}-b_{n-1}<z_{n})
=\det\Big(\rme^{-\frac{1}{2}\big(\sum\limits_{\ell=1}^{j}z_{\ell}-\sum\limits_{\ell=1}^{k}z_{\ell}\big)^{2}}\Big)_{j,k=0,\ldots,n}\;.
\end{equation}
The probability distribution (\ref{P bn stat}) implies in particular after some calculations
\begin{eqnarray}
\label{<max[b1-b0]max[b2-b1]>}
&& \langle\max[b_{1}-b_{0}]\max[b_{2}-b_{1}]\rangle_{b_{0},b_{1},b_{2}}=-\frac{1}{2}+\frac{2\pi}{3\sqrt{3}}\\
\label{<max[b1-b0]^2 max[b2-b1]>}
&& \langle\max[b_{1}-b_{0}]^{2}\max[b_{2}-b_{1}]\rangle_{b_{0},b_{1},b_{2}}=\frac{5\sqrt{\pi}}{12}\\
\label{<max[b1-b0] max[b2-b1]^2>}
&& \langle\max[b_{1}-b_{0}]\max[b_{2}-b_{1}]^{2}\rangle_{b_{0},b_{1},b_{2}}=\frac{5\sqrt{\pi}}{12}\;.
\end{eqnarray}
Furthermore, numerical integration gives
\begin{equation}
\label{<max[b1-b0]max[b2-b1]max[b3-b2]>}
\langle\max[b_{1}-b_{0}]\max[b_{2}-b_{1}]\max[b_{3}-b_{2}]\rangle_{b_{0},b_{1},b_{2},b_{3}}\approx0.565509\;.
\end{equation}
\end{subsubsection}

\begin{subsubsection}{Brownian averages for flat initial condition.}
\label{section elementary expectation values max flat}
\hfill\break
We compute in this section Brownian averages related to (\ref{theta flat[bb]}) with flat initial condition, which provide a partial check of conjecture \ref{conjecture flat}.

We consider independent standard Brownian bridges $b_{j}$ with $b_{j}(0)=b_{j}(1)=0$, $j=0,\ldots,n$. Then
\begin{eqnarray}
&&\fl\hspace{5mm} \P(b_{0}<z_{0},b_{1}-b_{0}<z_{1},\ldots,b_{n}-b_{n-1}<z_{n})\\
&&\fl\hspace{15mm} =\frac{\P(w_{0}<z_{0},w_{1}-w_{0}<z_{1},\ldots,w_{n}-w_{n-1}<z_{n},0<w_{0}(1)<\rmd y_{0},\ldots)}{\P(0<w_{0}(1)<\rmd y_{0},\ldots,0<w_{n}(1)<\rmd y_{n})}\;,\nonumber
\end{eqnarray}
where $w_{j}$, $j=0,\ldots,n$ are independent Wiener processes and all $\rmd y_{j}\to0$. The denominator is equal to $(\sqrt{2\pi})^{-n-1}\rmd y_{0}\ldots\rmd y_{n}$. The numerator is the distribution of a process of $n+1$ Brownian motions on the negative real axis that terminates when two particles collide or when one particle reaches the origin. Alternatively, the numerator can be thought of as the distribution of a $n+1$ dimensional Brownian motion in the Weyl chamber associated to root system $B_{n+1}$ with absorbing boundaries, and equation (9) of \cite{G1999.1} leads to
\begin{eqnarray}
\label{P bn flat}
&& \P(b_{0}<z_{0},b_{1}-b_{0}<z_{1},\ldots,b_{n}-b_{n-1}<z_{n})\\
&&\hspace{20mm} =\det\Big(\rme^{-\frac{1}{2}\big(\sum\limits_{\ell=0}^{j}z_{\ell}-\sum\limits_{\ell=0}^{k}z_{\ell}\big)^{2}}-\rme^{-\frac{1}{2}\big(\sum\limits_{\ell=0}^{j}z_{\ell}+\sum\limits_{\ell=0}^{k}z_{\ell}\big)^{2}}\Big)_{j,k=0,\ldots,n}\;.\nonumber
\end{eqnarray}
The probability distribution (\ref{P bn flat}) implies in particular after some calculations
\begin{eqnarray}
\label{<max[b0]max[b1-b0]>}
&& \langle\max[b_{0}]\max[b_{1}-b_{0}]\rangle_{b_{0},b_{1}}=-\frac{1}{2}+\frac{5\pi}{16}\\
\label{<max[b0]^2 max[b1-b0]>}
&& \langle\max[b_{0}]^{2}\max[b_{1}-b_{0}]\rangle_{b_{0},b_{1}}=\frac{13\sqrt{\pi}}{16}-\frac{7\sqrt{\pi}}{8\sqrt{2}}\\
\label{<max[b0] max[b1-b0]^2>}
&& \langle\max[b_{0}]\max[b_{1}-b_{0}]^{2}\rangle_{b_{0},b_{1}}=\frac{13\sqrt{\pi}}{8\sqrt{2}}-\frac{7\sqrt{\pi}}{8}\;.
\end{eqnarray}
Furthermore, numerical integration gives
\begin{equation}
\label{<max[b0]max[b1-b0]max[b2-b1]>}
\langle\max[b_{0}]\max[b_{1}-b_{0}]\max[b_{2}-b_{1}]\rangle_{b_{0},b_{1},b_{2}}\approx0.383683\;.
\end{equation}
\end{subsubsection}

\end{subsection}

\begin{subsection}{Perturbative expansion of $\langle\max[b-h]\rangle_{b}$ for $h$ of small amplitude}
\label{section <max[b-h]> small h}
We study in this section the perturbative expansion of $\langle\max[b-h]\rangle_{b}$ for $h$ of small amplitude, with $b(t)$ a standard Brownian bridge and $h(t)$ a continuous function with $h(0)=h(1)=0$. The variable is called $t$ in this section only in order to conform to usual notations for diffusive processes. The perturbative expansion is worked out independently from three approaches: a naive perturbative solution of the heat equation, the relation to first passage time, and the Cameron-Martin formula.

\begin{subsubsection}{Perturbative solution of the heat equation.}
\hfill\break
We consider in this section the probability density $P_{h}(t,y,z)$ defined as
\begin{equation}
\fl\hspace{15mm}
P_{h}(t,y,z)\rmd y=\P(\max_{0\leq s\leq t}(w(s)-h(s))<z,y<w(t)<y+\rmd y)\;,
\end{equation}
where $w(t)$, $w(0)=0$ is the Wiener process. The function $P_{h}$ is the solution of the heat equation with appropriate initial and boundary conditions:
\begin{eqnarray}
\label{Ph(t,y,z) heat eq}
&& \partial_{t}P_{h}(t,y,z)=\frac{1}{2}\partial_{y}^{2}P_{h}(t,y,z)\\
\label{Ph(0,y,z)}
&& P_{h}(0,y,z)=\delta(y)1_{\{y<z\}}\\
\label{Ph(t,z+h,z)}
&& P_{h}(t,z+h(t),z)=0\;.
\end{eqnarray}
In the special case $h=0$, the function $P_{0}(t,y,z)$ can be obtained directly from the method of images as
\begin{equation}
\label{P0(t,y,z)}
P_{0}(t,y,z)=\frac{\rme^{-\frac{y^{2}}{2t}}}{\sqrt{2\pi t}}-\frac{\rme^{-\frac{(2z-y)^{2}}{2t}}}{\sqrt{2\pi t}}\;,
\end{equation}
which is indeed the solution of (\ref{Ph(t,y,z) heat eq})-(\ref{Ph(t,z+h,z)}). We are interested in corrections to (\ref{P0(t,y,z)}) for $P_{\epsilon h}(t,y,z)$ when $\epsilon\to0$, and write
\begin{equation}
\label{Peh[Pjh]}
P_{\epsilon h}(t,y,z)=\sum_{j=0}^{\infty}P_{j,h}(t,y,z)\,\epsilon^{j}\;,
\end{equation}
with $P_{0,h}=P_{0}$. From (\ref{Ph(t,y,z) heat eq})-(\ref{Ph(t,z+h,z)}), the functions $P_{j,h}(t,y,z)$ are still solution of the heat equation, but with zero initial condition for $j\geq1$ and a boundary condition involving $P_{k,h}$, $k<j$:
\begin{eqnarray}
\label{Pj(t,y,z) heat eq}
&& \partial_{t}P_{j,h}(t,y,z)=\frac{1}{2}\partial_{y}^{2}P_{j,h}(t,y,z)\\
\label{Pj(0,y,z)}
&& P_{j,h}(0,y,z)=0\qquad(j\geq1)\\
\label{Pj^(0,k,0)(t,z,z) sum}
&& \sum_{k=0}^{j}\frac{h(t)^{k}}{k!}P_{j-k,h}^{(0,k,0)}(t,z,z)=0\;.
\end{eqnarray}
It is a classical result (see e.g. \cite{CJ1959.1}) that the solution of (\ref{Pj(t,y,z) heat eq}) with zero initial condition (\ref{Pj(0,y,z)}) can be expressed in terms of the space derivative of the heat kernel and the boundary value $P_{j}(t,z,z)$ as
\begin{equation}
\label{Pj(t,y,z)[Pj(t,z,z)]}
P_{j,h}(t,y,z)\underset{j\geq1}{=}\int_{0}^{t}\rmd s\,\frac{(z-y)\,\rme^{-\frac{(z-y)^{2}}{2(t-s)}}}{\sqrt{2\pi}\,(t-s)^{3/2}}P_{j,h}(s,z,z)\;.
\end{equation}
Inserting (\ref{Pj(t,y,z)[Pj(t,z,z)]}) into the boundary condition (\ref{Pj^(0,k,0)(t,z,z) sum}) for $k\neq0$ then gives a systematic recursive solution for $P_{j,h}(t,y,z)$. We find in particular
\begin{equation}
\label{P1h}
P_{1,h}(t,z,z)=\frac{2z\,\rme^{-\frac{z^{2}}{2t}}}{\sqrt{2\pi}\,t^{3/2}}\,h(t)
\end{equation}
and
\begin{equation}
\label{P2h}
\fl\hspace{5mm}
P_{2,h}(t,z,z)=-\frac{2z\,\rme^{-\frac{z^{2}}{2t}}}{\pi\,t^{2}}\,h(t)^{2}-\frac{zh(t)}{\pi}\int_{0}^{t}\frac{\rmd s}{(t-s)^{3/2}}\,\Big(\frac{\rme^{-\frac{z^{2}}{2t}}\,h(t)}{t^{3/2}}-\frac{\rme^{-\frac{z^{2}}{2s}}\,h(s)}{s^{3/2}}\Big)\;.
\end{equation}
Corresponding expressions for $P_{1,h}(t,y,z)$ and $P_{2,h}(t,y,z)$ are given by (\ref{Pj(t,y,z)[Pj(t,z,z)]}).

The statistics of $\max[b-h]=\max_{0\leq t\leq1}(b(t)-h(t))$ for $b$ a standard Brownian bridge (i.e. a Wiener process conditioned on the event $w(1)=0$) can be computed in terms of $P_{h}(1,0,z)$. Using $\P(\max[w-h]<z|w(1)=0)=\lim_{\rmd y\to0}\P(\max[w-h]<z,0<w(1)<\rmd y)/\P(0<w(1)<\rmd y)$ and $\P(0<w(1)<\rmd y)=\rmd y/\sqrt{2\pi}$, one has
\begin{equation}
\label{P(max[b-h]<z)[Ph(1,0,z)]}
\P(\max[b-h]<z)=\sqrt{2\pi}\,P_{h}(1,0,z)\;.
\end{equation}
For small $\epsilon$, we write $\langle\max[b-\epsilon h]\rangle_{b}=\mu_{0}[h]+\mu_{1}[h]\,\epsilon+\mu_{2}[h]\,\epsilon^{2}+\ldots$. Then, (\ref{P(max[b-h]<z)[Ph(1,0,z)]}) supplemented by the perturbative expansion (\ref{Peh[Pjh]}) with (\ref{P0(t,y,z)}), (\ref{P1h}) and (\ref{P2h}) gives at first orders in $\epsilon$
\begin{eqnarray}
\label{mu0h}
&& \mu_{0}[h]=\frac{\sqrt{2\pi}}{4}\\
\label{mu1h}
&& \mu_{1}[h]=-\int_{0}^{1}\rmd t\,h(t)\\
\label{mu2h}
&& \mu_{2}[h]=\int_{0}^{1}\rmd t\,
\Bigg(\frac{2h(t)^{2}}{\sqrt{2\pi t}}-\int_{0}^{t}\rmd s\,\Big(\frac{h(s)h(t)}{\sqrt{2\pi}(t-s)^{3/2}(1+s-t)^{3/2}}\\
&&\hspace{85mm} -\frac{h(t)^{2}}{\sqrt{2\pi}(t-s)^{3/2}}\Big)\Bigg)\;.\nonumber
\end{eqnarray}
The first coefficient (\ref{mu0h}) is the well known expectation value of the maximum of a standard Brownian bridge.

It turns out that the first $\mu_{n}[h]$ have much simpler expressions in terms of the Fourier coefficients $a_{k}$ of $h$ defined as
\begin{equation}
h(t)=\sum_{k\in\mathbb{Z}}a_{k}\,\rme^{2\rmi\pi kt}\;.
\end{equation}
One has $\mu_{1}[h]=-a_{0}$, and after some calculations, the double integrals for the coefficient $\mu_{2}[h]$ can be rewritten as
\begin{equation}
\label{mu2h[a]}
\mu_{2}[h]=-\sqrt{2\pi}\,\sum_{k\in\mathbb{Z}}a_{k}a_{-k}(-1)^{k}k\pi J_{1}(k\pi)
\end{equation}
with $J$ the Bessel function of the first kind. This proves equation (\ref{max[b-h] small h}).
\end{subsubsection}

\begin{subsubsection}{First passage time to the boundary.}
\hfill\break
For a Wiener process $w(t)$, the first passage time $T[w]$ to the boundary $h(t)+z$ with $z>0$ is the smallest $t>0$ such that $w(t)=h(t)+z$. The probability density of $T[w]$ is written as $p_{\text{w}}$ in the following, with the dependency in $h$ and $z$ kept implicit. Similarly, we write $p_{\bb}$ for the probability density of the first passage time of a standard Brownian bridge to the boundary $h(t)+z$. Using the fact that $b(t)$ is equal in law to $w(t)$ conditioned on $w(1)=0$ and the Markov property of the Wiener process, the densities $p_{\text{w}}$ and $p_{\bb}$ are related for $t<1$ by
\begin{equation}
\label{first passage density bb}
p_{\bb}(t)=\sqrt{2\pi}\,p_{\text{w}}(t)\,\frac{\rme^{-\frac{(h(t)+z)^{2}}{2(1-t)}}}{\sqrt{2\pi(1-t)}}\;.
\end{equation}
The density $p_{\text{w}}$ was first computed explicitly by Durbin \cite{DW1992.1}, see also \cite{P2002.1}, as
\begin{eqnarray}
\label{first passage density w}
&&
\fl\hspace{5mm}
p_{\text{w}}(t_{0})=\sum_{j=1}^{\infty}(-1)^{j}\int_{0<t_{j-1}<\ldots<t_{1}<t_{0}}\rmd t_{1}\ldots\rmd t_{j-1}\,\Big(\frac{h(t_{j-1})+z}{t_{j-1}}-h'(t_{j-1})\Big)\\
&&
\fl\hspace{35mm}
\times\prod_{i=1}^{j-1}\Big(\frac{h(t_{i-1})-h(t_{i})}{t_{i-1}-t_{i}}-h'(t_{i-1})\Big)\,\frac{\rme^{-\frac{(h(t_{j-1})+z)^{2}}{2t_{j-1}}}}{\sqrt{2\pi t_{j-1}}}\prod_{i=1}^{j-1}\frac{\rme^{-\frac{(h(t_{i-1})-h(t_{i}))^{2}}{2(t_{i-1}-t_{i})}}}{\sqrt{2\pi(t_{i-1}-t_{i})}}\;.\nonumber
\end{eqnarray}
This expression for $p_{\text{w}}$ can be derived from an integral equation resulting from the conservation of probability (Chapman-Kolmogorov equation) with the boundary as the intermediate point. Using $\P(\max[b-h]<z)=1-\int_{0}^{1}\rmd t\,p_{\bb}(t)$, the average $\langle\max[b-h]\rangle_{b}$ can finally be expressed in terms of the first passage density $p_{\bb}$ as
\begin{equation}
\langle\max[b-h]\rangle_{b}=-\int_{0}^{\infty}\rmd z\,z\partial_{z}\int_{0}^{1}\rmd t\,p_{\bb}(t)\;.
\end{equation}
Expanding (\ref{first passage density bb}) and (\ref{first passage density w}) for $h$ of small amplitude, the integral over $z$ can be performed explicitly, and we recover (\ref{mu0h}), (\ref{mu1h}) after some calculations.
\end{subsubsection}

\begin{subsubsection}{Cameron-Martin formula.}
\hfill\break
We consider in this section the Cameron-Martin formula \cite{CM1944.1}, which describes the change of the Wiener measure under translations. Let $\mathcal{F}$ be a functional acting on continuous functions on the interval $[0,1]$ and $h$ a continuous function on the same interval with $h(0)=0$. Under technical hypotheses on $\mathcal{F}$ and $h$, one has $\langle\mathcal{F}[w-h]\rangle_{w}=\rme^{-\frac{1}{2}\int_{0}^{1}\rmd t\,h'(t)^{2}}\langle\mathcal{F}[w]\,\rme^{-\int_{0}^{1}\rmd t\,h'(t)w'(t)}\rangle_{w}$, where the expectations values are computed with respect to the Wiener process $w(t)$, $w(0)=0$. Choosing a function $h$ that verifies additionally $h(1)=0$ and replacing $\mathcal{F}[g]$ by $\mathcal{F}[g]\openone_{\{g(1)=0\}}$, this implies the identity
\begin{equation}
\label{Cameron-Martin bb}
\langle\mathcal{F}[b-h]\rangle_{b}=\rme^{-\frac{1}{2}\int_{0}^{1}\rmd t\,h'(t)^{2}}\langle\mathcal{F}[b]\,\rme^{-\int_{0}^{1}\rmd t\,h'(t)b'(t)}\rangle_{b}\;,
\end{equation}
where $b$ is the standard Brownian bridge. Assuming $h'$ is continuous and taking $\mathcal{F}=\max$, one finds
\begin{eqnarray}
&& \langle\max[b-h]\rangle_{b}=\rme^{-\frac{1}{2}\int_{0}^{1}\rmd t\,h'(t)^{2}}\sum_{n=0}^{\infty}\frac{1}{n!}\int_{0}^{1}\rmd t_{1}\ldots\rmd t_{n}\,h''(t_{1})\ldots h''(t_{n})\\
&&\hspace{80mm} \times\langle b(t_{1})\ldots b(t_{n})\max[b]\rangle_{b}\;.\nonumber
\end{eqnarray}
The remaining expectation value can be computed from
\begin{eqnarray}
&&
\fl\hspace{5mm}
\P(\max b<z,y_{1}<b(t_{1})<y_{1}+\rmd y_{1},\ldots,y_{n}<b(t_{n})<y_{n}+\rmd y_{n})\\
&&
\fl\hspace{10mm}
=\rmd y_{1}\ldots\rmd y_{n}\,\sqrt{2\pi}\Big(\frac{\rme^{-\frac{y_{1}^{2}}{2t_{1}}}}{\sqrt{2\pi t_{1}}}-\frac{\rme^{-\frac{(2z-y_{1})^{2}}{2t_{1}}}}{\sqrt{2\pi t_{1}}}\Big)
\Big(\frac{\rme^{-\frac{(y_{2}-y_{1})^{2}}{2(t_{2}-t_{1})}}}{\sqrt{2\pi(t_{2}-t_{1})}}-\frac{\rme^{-\frac{(2z-y_{1}-y_{2})^{2}}{2(t_{2}-t_{1})}}}{\sqrt{2\pi(t_{2}-t_{1})}}\Big)\times\ldots\nonumber\\
&&
\fl\hspace{15mm}
\ldots\times\Big(\frac{\rme^{-\frac{(y_{n}-y_{n-1})^{2}}{2(t_{n}-t_{n-1})}}}{\sqrt{2\pi (t_{n}-t_{n-1})}}-\frac{\rme^{-\frac{(2z-y_{n-1}-y_{n})^{2}}{2(t_{n}-t_{n-1})}}}{\sqrt{2\pi(t_{n}-t_{n-1})}}\Big)
\Big(\frac{\rme^{-\frac{y_{n}^{2}}{2(1-t_{n})}}}{\sqrt{2\pi(1-t_{n})}}-\frac{\rme^{-\frac{(2z-y_{n})^{2}}{2(1-t_{n})}}}{\sqrt{2\pi(1-t_{n})}}\Big)\;,\nonumber
\end{eqnarray}
valid in the sector $0<t_{1}<\ldots<t_{n}<1$, and which follows directly from (\ref{P0(t,y,z)}). Expanding for $h$ with small amplitude, the identity
\begin{eqnarray}
&&
\fl\hspace{5mm}
\sqrt{2\pi}\int_{0}^{\infty}\rmd z\int_{-\infty}^{z}\rmd y\,yz\partial_{z}
\Big(\frac{\rme^{-\frac{y^{2}}{2t}}}{\sqrt{2\pi t}}-\frac{\rme^{-\frac{(2z-y)^{2}}{2t}}}{\sqrt{2\pi t}}\Big)
\Big(\frac{\rme^{-\frac{y^{2}}{2(1-t)}}}{\sqrt{2\pi(1-t)}}-\frac{\rme^{-\frac{(2z-y)^{2}}{2(1-t)}}}{\sqrt{2\pi(1-t)}}\Big)\nonumber\\
&&
\fl\hspace{5mm}
=\frac{t(1-t)}{2}
\end{eqnarray}
allows to recover $\max[b-h]\simeq\frac{\sqrt{2\pi}}{4}-\int_{0}^{1}\rmd t\,h(t)$.
\end{subsubsection}
\end{subsection}

\begin{subsection}{$\langle\max[b-h]\rangle_{b}$ for $h$ piecewise linear}
\label{section piecewise linear}
In all this section, we restrict to the special case $h(x)=cx\openone_{\{0\leq x<a\}}+ca\,\frac{1-x}{1-a}\openone_{\{a\leq x<1\}}$ with $0<a<1$, corresponding for TASEP to a small initial domain wall with reduced density profile $\sigma_{0}(x)=-c\openone_{\{0\leq x<a\}}+\frac{ca}{1-a}\openone_{\{a\leq x<1\}}$. We compute $\langle\max[b-h]\rangle_{b}$ using the Cameron-Martin formula, see figure \ref{fig piecewise and parabola} for a plot as a function of the amplitude $c$. Perturbative expansions for small and large $c$ are also studied.

\begin{subsubsection}{Exact formula.}
\hfill\break
We consider the Cameron-Martin formula (\ref{Cameron-Martin bb}) with the functional $\mathcal{F}=\max$. After splitting the integrals at $x=a$ since $h'(x)$ is not continuous at that point, partial integration of the integral inside the expectation value on the right leads to
\begin{equation}
\label{max[b-h][exp(...)max[b]] piecewise linear}
\langle\max[b-h]\rangle_{b}=\rme^{-\frac{ac^{2}}{2(1-a)}}\big\langle\rme^{-\frac{c}{1-a}\,b(a)}\max[b]\big\rangle_{b}\;.
\end{equation}
The joint probability of $\max[b]$ and $b(a)$ is needed in order to evaluate the expectation value on the right hand side. Introducing the notation $w(a)\in\rmd y$ for $w(a)\in(y,y+\rmd y)$, the definition of the standard Brownian bridge $b(x)$ as a Wiener process $w(x)$ conditioned on $w(1)=0$, followed by the Markov property of the Wiener process, leads to
\begin{eqnarray}
&&\hspace{-5mm} \P(\max[b]<z,b(a)\in\rmd y)\nonumber\\
&&\hspace{-5mm} =\frac{\P(\max\limits_{0<x<a}w(x)<z,w(a)\in\rmd y,\max\limits_{a<x<1}w(x)<z,w(1)\in(0,\rmd u)|w(0)=0)}{\P(w(1)\in\rmd u|w(0)=0)}\nonumber\\
&&\hspace{-5mm} =\frac{\sqrt{2\pi}}{\rmd u}\,\P(\max\limits_{0<x<a}w(x)<z,w(a)\in\rmd y|w(0)=0)\\
&&\hspace{10mm} \times\P(\max\limits_{a<x<1}w(x)<z,w(1)\in(0,\rmd u)|w(a)=y)\;.\nonumber
\end{eqnarray}
Using translation invariance of the Wiener process, the two remaining probabilities are given by (\ref{P0(t,y,z)}). Inserting into (\ref{max[b-h][exp(...)max[b]] piecewise linear}), we obtain
\begin{eqnarray}
\label{max[b-h] piecewise linear}
&& \langle\max[b-h]\rangle_{b}=\frac{\rme^{-\frac{ac^{2}}{2(1-a)}}}{\sqrt{2\pi a(1-a)}}\,\int_{0}^{\infty}\rmd z\,z\,\partial_{z}\int_{0}^{\infty}\rmd y\,\rme^{-\frac{c(z-y)}{1-a}}\\
&&\hspace{45mm} \times\Big(\rme^{-\frac{(z-y)^{2}}{2a}}-\rme^{-\frac{(z+y)^{2}}{2a}}\Big)\Big(\rme^{-\frac{(z-y)^{2}}{2(1-a)}}-\rme^{-\frac{(z+y)^{2}}{2(1-a)}}\Big)\;.\nonumber
\end{eqnarray}
The integral over $y$ can be computed in terms of the error function $\erf$ as
\begin{eqnarray}
\label{P(max[b-h]<z) h linear erf}
&&
\fl\hspace{5mm}
\langle\max[b-h]\rangle_{b}=\int_{0}^{\infty}\rmd z\,z\partial_{z}\Big[\frac{1}{2}\Big(1+\erf\Big(\frac{ca+z}{\sqrt{2a(1-a)}}\Big)\Big)\\
&&
\fl\hspace{10mm}
-\frac{\rme^{-2cz-2z^{2}}}{2}\Big(1+\erf\Big(\frac{ca-z+2az}{\sqrt{2a(1-a)}}\Big)\Big)
-\frac{\rme^{-\frac{2acz}{1-a}-2z^{2}}}{2}\Big(1+\erf\Big(\frac{ca+z-2az}{\sqrt{2a(1-a)}}\Big)\Big)\nonumber\\
&&\fl\hspace{90mm}
+\frac{\rme^{-\frac{2cz}{1-a}}}{2}\Big(1+\erf\Big(\frac{ca-z}{\sqrt{2a(1-a)}}\Big)\Big)\Big]\;.\nonumber
\end{eqnarray}
Furthermore, in the symmetric case $a=1/2$, both integrals of (\ref{max[b-h] piecewise linear}) can be computed explicitly, and one has
\begin{eqnarray}
\label{max[b-h] piecewise linear symmetric}
&& \langle\max[b-h]\rangle_{b}=-\frac{1}{4c}-\frac{c-c^{-1}}{4}\Big(1-\erf\Big(\frac{c}{\sqrt{2}}\Big)\Big)\\
&&\hspace{32mm} +\frac{\rme^{-\frac{c^{2}}{2}}}{2\sqrt{2\pi}}+\frac{\pi\,\rme^{\frac{c^{2}}{2}}}{2\sqrt{2\pi}}\Big(1+\erf\Big(\frac{c}{\sqrt{2}}\Big)\Big)\Big(1-\erf\Big(\frac{c}{\sqrt{2}}\Big)\Big)\;.\nonumber
\end{eqnarray}
\end{subsubsection}

\begin{subsubsection}{Small $c$ perturbative expansion.}
\hfill\break
From (\ref{max[b-h] piecewise linear}), the perturbative expansion near $c=0$ of $\langle\max[b-h]\rangle_{b}$ gives for arbitrary $a$
\begin{eqnarray}
\label{max[b-h] piecewise linear small c}
&&
\fl\hspace{2mm}
\langle\max[b-h]\rangle_{b}=\frac{\sqrt{2\pi}}{4}-\frac{ac}{2}+\Big(\frac{\sqrt{2\pi}}{16}-\frac{\sqrt{a(1-a)}}{2\sqrt{2\pi}}-\frac{a(1-a)\sqrt{2\pi}}{8}\\
&&
\fl\hspace{7mm}
+\frac{2a^{3/2}(1-a)^{3/2}}{3\sqrt{2\pi}}+\frac{1-2a}{4\sqrt{2\pi}}\Big(\arctan\sqrt{\frac{a}{1-a}}-\arctan\sqrt{\frac{1-a}{a}}\Big)\Big)\frac{c^{2}}{(1-a)^{2}}+\mathcal{O}(c^{3})\;.\nonumber
\end{eqnarray}
\hfill\break
{\bf Remark:} Comparing the terms of order two in $c$ in the above formula with (\ref{max[b-h] small h}) and using the Fourier series representation $h(x)=\sum_{k=-\infty}^{\infty}a_{k}\rme^{2\rmi\pi kx}$ with $a_{0}=ac/2$ and $a_{k}=\frac{c}{1-a}\,\frac{\rme^{-2\rmi\pi ka}-1}{4\pi^{2}k^{2}}$, $k\neq0$, we find the following identity valid for arbitrary values of $a$:
\begin{eqnarray}
\label{sum Bessel piecewise linear}
&&\fl\hspace{1mm} \sum_{k=1}^{\infty}\frac{(-1)^{k}J_{1}(k\pi)\sin^{2}(\pi ka)}{\pi^{2}k^{3}}=-\frac{\pi}{8}+\frac{\sqrt{a(1-a)}}{2}+\frac{\pi a(1-a)}{4}-\frac{2a^{3/2}(1-a)^{3/2}}{3}\\
&&\hspace{36mm} -\frac{1-2a}{4}\Big(\arctan\sqrt{\frac{a}{1-a}}-\arctan\sqrt{\frac{1-a}{a}}\Big)\;.\nonumber
\end{eqnarray}
We have not been able to find this identity in the literature.
\end{subsubsection}

\begin{subsubsection}{Asymptotics for $c\to+\infty$.}
\hfill\break
After making the change of variable $z=w/c$ in (\ref{P(max[b-h]<z) h linear erf}), the large $c$ limit of all four error functions is $\erf(\ldots)\simeq1$ up to exponentially small terms. Expanding the remaining factors at large $c$ and computing the integral over $w$ finally gives the asymptotic expansion
\begin{equation}
\label{max[b-h] piecewise linear c>>1}
\langle\max[b-h]\rangle_{b}\simeq\frac{1-a+a^{2}}{2ac}+\sum_{k=1}^{\infty}\frac{(-1)^{k}(2k)!}{2^{k+1}k!c^{2k+1}}\Big(1+\Big(\frac{1-a}{a}\Big)^{2k+1}\Big)\;.
\end{equation}
For the symmetric case $a=1/2$, the leading term is in particular $\langle\max[b-h]\rangle_{b}\simeq\frac{3}{4c}$, in agreement with $\langle\max_{x\in\mathbb{R}}(w(x)-|x|)\rangle_{w}=3/4$ for a two-sided Wiener process, as explained in section \ref{section <h>} for $h_{0}$ of large amplitude.
\end{subsubsection}

\begin{subsubsection}{Asymptotics for $c\to-\infty$.}
\hfill\break
We start again with (\ref{P(max[b-h]<z) h linear erf}), make the change of variable $z=w-ca$ and use $1+\erf(q)\simeq-\frac{\rme^{-q^{2}}}{\sqrt{\pi}}\sum_{k=0}^{\infty}\frac{(-1)^{k}(2k)!}{2^{2k}k!q^{2k+1}}$ when $q\to-\infty$ on the three error functions with large argument. Neglecting exponentially small terms, the integral over $w\in(ca,\infty)$ in $\langle\max[b-h]\rangle_{b}$ can be extended to $w\in\mathbb{R}$. After further large $|c|$ expansions, the Gaussian integrals over $w$ can be computed. A little algebra finally gives
\begin{equation}
\label{max[b-h] piecewise linear c<<-1}
\fl\hspace{10mm} \langle\max[b-h]\rangle_{b}\simeq-ca-\frac{1-a+a^{2}}{2ac}-\sum_{k=1}^{\infty}\frac{(-1)^{k}(2k)!}{2^{k+1}k!c^{2k+1}}\Big(1+\Big(\frac{1-a}{a}\Big)^{2k+1}\Big)\;.
\end{equation}
This is essentially the same asymptotic series as when $c\to+\infty$. The extra leading term $-ca$ is simply equal to $-\min[h]$ when $c<0$, in agreement with the discussion in section \ref{section <h>} for $h_{0}$ of large amplitude.
\end{subsubsection}
\end{subsection}

\begin{subsection}{$\langle\max[b-h]\rangle_{b}$ for the parabola}
In all this section, we restrict to the special case $h(x)=cx(1-x)$, corresponding for TASEP to an initial linear ramp with reduced density profile $\sigma_{0}(x)=2c(x-1/2)$. We write an expression for $\langle\max[b-h]\rangle_{b}$ using exact results \cite{G1989.1} for the Wiener process absorbed by a parabola, see figure \ref{fig piecewise and parabola} for a plot as a function of the amplitude $c$. Perturbative expansions for small and large $c$ are also studied.

\begin{subsubsection}{Exact formula.}
\hfill\break
From \cite{G1989.1}, the probability that a standard Wiener process $w(x)$ with $w(0)=0$ stays under a parabola is known explicitly. One has for $c<0$
\begin{equation}
\P\Big(\max_{0\leq x\leq1}(w(x)-cx(1-x))<z, 0<w(1)<\rmd y\Big)=K_{z,c}(1)\,\rmd y\;,
\end{equation}
where the Laplace transform of $K_{z,c}$ is
\begin{equation}
\fl\hspace{5mm}
\int_{0}^{\infty}\rmd u\,\rme^{-\lambda u}K_{z,c}(u)=\frac{2\pi\rme^{-2cz-\frac{c^{2}}{6}}}{|4c|^{1/3}}\,\Ai^{2}\Big(\frac{2\lambda-4cz}{|4c|^{2/3}}\Big)\Bigg(\frac{\Bi\Big(\frac{2\lambda-4cz}{|4c|^{2/3}}\Big)}{\Ai\Big(\frac{2\lambda-4cz}{|4c|^{2/3}}\Big)}-\frac{\Bi\Big(\frac{2\lambda}{|4c|^{2/3}}\Big)}{\Ai\Big(\frac{2\lambda}{|4c|^{2/3}}\Big)}\Bigg)\;,
\end{equation}
with $\Ai$ and $\Bi$ the Airy functions (see e.g. \cite{DLMF} section 9). Since the zeroes $a_{j}$ of $\Ai$ are on the negative real axis, the Laplace transform can be inverted with an integral on the imaginary axis:
\begin{equation}
\fl\hspace{5mm}
K_{z,c}(u)=\frac{\rme^{-2cz-\frac{c^{2}}{6}}}{\rmi|4c|^{1/3}}\int_{-\rmi\infty}^{\rmi\infty}\rmd\lambda\,\rme^{\lambda u}\Ai^{2}\Big(\frac{2\lambda-4cz}{|4c|^{2/3}}\Big)\Bigg(\frac{\Bi\Big(\frac{2\lambda-4cz}{|4c|^{2/3}}\Big)}{\Ai\Big(\frac{2\lambda-4cz}{|4c|^{2/3}}\Big)}-\frac{\Bi\Big(\frac{2\lambda}{|4c|^{2/3}}\Big)}{\Ai\Big(\frac{2\lambda}{|4c|^{2/3}}\Big)}\Bigg)\;.
\end{equation}
Then, using the relation $\partial_{y}\frac{\Bi(y)}{\Ai(y)}=\frac{1}{\pi\Ai^{2}(y)}$, one finds after a few changes of variables and a shift of the contour for $\lambda$
\begin{equation}
K_{z,c}(1)=\frac{\rme^{-\frac{c^{2}}{6}}}{\rmi\pi}\int_{0}^{z}\rmd y\int_{-\rmi\infty}^{\rmi\infty}\rmd\lambda\,\rme^{\lambda-2cy}\frac{\Ai^{2}(\frac{2\lambda-4cy}{|4c|^{2/3}})}{\Ai^{2}(\frac{2\lambda}{|4c|^{2/3}})}\;.
\end{equation}
We obtain for the standard Brownian bridge $b(x)$ and the parabola $h(x)=cx(1-x)$ with $c<0$
\begin{equation}
\label{max[b-h] parabola int lambda}
\langle\max[b-h]\rangle_{b}=\sqrt{2\pi}\,\frac{\rme^{-\frac{c^{2}}{6}}}{\rmi\pi}\int_{0}^{\infty}\rmd z\int_{-\rmi\infty}^{\rmi\infty}\rmd\lambda\,z\,\rme^{\lambda-2cz}\frac{\Ai^{2}(\frac{2\lambda-4cz}{|4c|^{2/3}})}{\Ai^{2}(\frac{2\lambda}{|4c|^{2/3}})}\;.
\end{equation}
Alternatively, the integral over $\lambda$ can be computed from residues at the zeroes of $\Ai^{2}(\frac{2\lambda}{|4c|^{2/3}})$. Taking into account only the $J$ first zeroes $a_{j}$ for the moment, we find after partial integration in the variable $z$
\begin{eqnarray}
\label{max[b-h] parabola int lambda -> sum j}
&&
\fl\hspace{10mm}
\langle\max[b-h]\rangle_{b}=\sqrt{2\pi}\,\frac{\rme^{-\frac{c^{2}}{6}}}{\rmi\pi}\int_{0}^{\infty}\rmd z\int_{-\rmi\infty+\frac{|4c|^{2/3}}{2}\frac{a_{J}+a_{J+1}}{2}}^{\rmi\infty+\frac{|4c|^{2/3}}{2}\frac{a_{J}+a_{J+1}}{2}}\rmd\lambda\,z\,\rme^{\lambda-2cz}\frac{\Ai^{2}(\frac{2\lambda-4cz}{|4c|^{2/3}})}{\Ai^{2}(\frac{2\lambda}{|4c|^{2/3}})}\nonumber\\
&&
\fl\hspace{40mm}
\hspace{30mm}
-\sqrt{2\pi}\,\rme^{-\frac{c^{2}}{6}}\sum_{j=1}^{J}\int_{a_{j}}^{\infty}\!\!\rmd z\,\rme^{\frac{|4c|^{2/3}z}{2}}\frac{\Ai(z)^{2}}{\Ai'(a_{j})^{2}}\;.
\end{eqnarray}
Using the derivative with respect to $q$ of the relation $\int_{-\infty}^{\infty}\rmd z\,\rme^{qz}\Ai^{2}(z+a)=\frac{\rme^{\frac{q^{3}}{12}-qa}}{\sqrt{4\pi q}}$, we express the integral for $z$ between $0$ and infinity in (\ref{max[b-h] parabola int lambda -> sum j}) as an explicit term proportional to $\rme^{c^{2}/6}$ minus an integral for $z$ between $-\infty$ and $0$, which vanishes when $J\to\infty$. The remaining integral over $\lambda$ can then be computed at large $J$ from $\int_{-\rmi\infty}^{\rmi\infty}\frac{\lambda\,\rmd\lambda}{\cos^{2}(\lambda)}=0$ and $\int_{-\rmi\infty}^{\rmi\infty}\frac{\rmd\lambda}{\cos^{2}(\lambda)}=2\rmi$ after using the asymptotics $\Ai(x)\simeq\frac{\cos(\frac{2}{3}(-x)^{3/2}-\pi/4)}{\sqrt{\pi}(-x)^{1/4}}$ for $x$ close to the negative real axis. We finally find the alternative expression
\begin{equation}
\label{max[b-h] parabola sum j}
\fl
\langle\max[b-h]\rangle_{b}=\lim_{J\to\infty}\frac{(3\pi J/2)^{2/3}}{|4c|^{1/3}}-\frac{c}{4}+\frac{1}{4c}-\sqrt{2\pi}\,\rme^{-\frac{c^{2}}{6}}\sum_{j=1}^{J}\int_{a_{j}}^{\infty}\!\!\rmd z\,\rme^{\frac{|4c|^{2/3}z}{2}}\frac{\Ai(z)^{2}}{\Ai'(a_{j})^{2}}\;.
\end{equation}
\end{subsubsection}

\begin{subsubsection}{Large $c$ asymptotics, $c<0$.}
\hfill\break
The large $|c|$ expansion of $\langle\max[b-h]\rangle_{b}$ can be extracted from (\ref{max[b-h] parabola int lambda}). After making the changes of variables $\lambda=|4c|^{2/3}\mu/2$ and $z=y-c/4-\mu/|4c|^{1/3}$, we observe that the exponentially large factors of the integrand cancel since $\Ai(x)\simeq\frac{\rme^{-2x^{3/2}/3}}{2\sqrt{\pi}x^{1/4}}$ at large $x$ away from the negative real axis. In particular, we find that the integrand is proportional to $\rme^{-2y^{2}}$. Since the integral over $y$ is between $c/4+\mu/|4c|^{1/3}$ and $+\infty$, the missing part of the integral between $-\infty$ and $c/4+\mu/|4c|^{1/3}$ gives an exponentially small contribution to $\langle\max[b-h]\rangle_{b}$. Neglecting these exponentially small terms, the integrals over $\mu$ give two constants:
\begin{equation}
\label{Xi parabola int lambda}
\int_{-\rmi\infty}^{\rmi\infty}\frac{\rmd\mu}{2\rmi\pi}\,\frac{1}{\Ai(\mu)^{2}}=1
\qquad\text{and}\qquad
\int_{-\rmi\infty}^{\rmi\infty}\frac{\rmd\mu}{2\rmi\pi}\,\frac{-\mu}{\Ai(\mu)^{2}}=\Xi\approx1.25512\;.
\end{equation}
The remaining integral over $y$ can be computed using $\int_{-\infty}^{\infty}\rmd z\,\rme^{qz}\Ai^{2}(z+a)=\frac{\rme^{\frac{q^{3}}{12}-qa}}{\sqrt{4\pi q}}$ and its derivative with respect to $q$. In the end, we find
\begin{equation}
\label{max[b-h] parabola c<<-1}
\langle\max[b-h]\rangle_{b}\simeq-\frac{c}{4}+\frac{\Xi}{|4c|^{1/3}}+\frac{1}{4c}
\end{equation}
up to exponentially small terms signaling the presence of an essential singularity when $c\to-\infty$.

The asymptotics (\ref{max[b-h] parabola c<<-1}) can also be recovered from (\ref{max[b-h] parabola sum j}), with a better characterization of the exponentially small terms, by using again $\int_{-\infty}^{\infty}\rmd z\,\rme^{qz}\Ai^{2}(z+a)=\frac{\rme^{\frac{q^{3}}{12}-qa}}{\sqrt{4\pi q}}$ to rewrite (\ref{max[b-h] parabola sum j}) with an integral over $z$ between $-\infty$ and $a_{j}$. We find at leading orders in $|c|$
\begin{equation}
\langle\max[b-h]\rangle_{b}\simeq-\frac{c}{4}+\frac{\Xi}{|4c|^{1/3}}+\frac{1}{4c}+\frac{\sqrt{2\pi}}{c^{2}}\,\rme^{-\frac{c^{2}}{6}+\frac{|4c|^{2/3}a_{1}}{2}}\;,
\end{equation}
with an alternative expression for the constant $\Xi$,
\begin{equation}
\label{Xi parabola sum j}
\Xi=\lim_{J\to\infty}\Big(\frac{3\pi J}{2}\Big)^{2/3}-\sum_{j=1}^{J}\frac{1}{\Ai'(a_{j})^{2}}\;.
\end{equation}

As explained in section \ref{section <h>}, the constant $\Xi$ is universal for all functions $h$ quadratic around their global minimum, if this minimum is reached only once.
\end{subsubsection}

\begin{subsubsection}{Large $c$ asymptotics, $c>0$.}
\hfill\break
The analytic continuation to $c>0$ of (\ref{max[b-h] parabola int lambda}) would be needed. We turn instead to Bethe ansatz numerics, which indicate that
\begin{equation}
\label{max[b-h] parabola c>>1}
\langle\max[b-h]\rangle_{b}\simeq\frac{3}{4c}
\end{equation}
when $c\to+\infty$. This is expected from the discussion about $h_{0}$ of large amplitude in section \ref{section <h>} since when $c>0$, $h(x)$ behaves near the location $x=0$ of its global minimum as $h(x)\simeq|x|$.
\end{subsubsection}

\begin{subsubsection}{Small $c$ perturbative expansion.}
\hfill\break
We start with (\ref{max[b-h] parabola int lambda}) and shift the contour for $\lambda$ to the line $r+\rmi\mathbb{R}$ with $r>c^{2}/2>0$. When $c\to0$, the arguments of the Airy functions become large and stay away from the negative real axis. We can then use the asymptotics \cite{DLMF}
\begin{equation}
\label{Ai asymptotics}
\Ai(x)\simeq\frac{\rme^{-\frac{2}{3}\,x^{3/2}}}{2\sqrt{\pi}\,x^{1/4}}\sum_{k=0}^{\infty}\frac{(6k-1)!!\,(-1)^{k}}{(2k-1)!!\,k!\,144^{k}x^{3k/2}}\;.
\end{equation}
After the small $c$ expansion, the integrand of (\ref{max[b-h] parabola int lambda}) has the form
\begin{equation}
\fl\hspace{5mm}
\sqrt{2\pi}\,\frac{\rme^{-\frac{c^{2}}{6}}}{\rmi\pi}\,z\,\rme^{\lambda-2cz}\frac{\Ai^{2}(\frac{2\lambda-4cz}{|4c|^{2/3}})}{\Ai^{2}(\frac{2\lambda}{|4c|^{2/3}})}
\simeq\frac{\rme^{-\frac{c^{2}}{6}}}{\sqrt{2\pi}\,\rmi}\,\rme^{\lambda-2z\sqrt{2\lambda}-2cz}\Big(2z+\sum_{k=0}^{\infty}c^{k}\sum_{j=0}^{2k}\frac{b_{j,k}z^{j+1}}{(2\lambda)^{\frac{3k-j}{2}}}\Big)\;,
\end{equation}
where the constants $b_{j,k}$ are rational numbers. Since $r>c^{2}/2$, the integral over $z$ converges, and one has
\begin{equation}
\fl\hspace{5mm}
\langle\max[b-h]\rangle_{b}=\frac{\rme^{-\frac{c^{2}}{6}}}{\sqrt{2\pi}\,\rmi}\sum_{k=0}^{\infty}c^{k}\sum_{j=0}^{2k}b_{j,k}\frac{(j+1)!}{2^{j+2}}\int_{r-\rmi\infty}^{r+\rmi\infty}\rmd\lambda\,\frac{\rme^{\lambda}}{(2\lambda)^{\frac{3k}{2}+1}(1-\frac{c}{\sqrt{2\lambda}})^{j+2}}\;.
\end{equation}
The remaining integral over $\lambda$ can be computed after expanding the integrand near $c=0$ using $\int_{r-\rmi\infty}^{r+\rmi\infty}\rmd\lambda\,\rme^{\lambda}/\lambda^{m+1}=2\rmi\pi/m!$ and $\int_{r-\rmi\infty}^{r+\rmi\infty}\rmd\lambda\,\rme^{\lambda}/\lambda^{m+1/2}=2^{m+1}\rmi\sqrt{\pi}/(2m-1)!!$ for $m$ non-negative integer. In the end, we obtain the perturbative expansion
\begin{eqnarray}
\label{max[b-h] parabola small c}
&&
\langle\max[b-h]\rangle_{b}=
\frac{\sqrt{2\pi}}{4}
-\frac{c}{6}
+\frac{\sqrt{2\pi}}{192}\,c^{2}
+\frac{1}{945}\,c^{3}
-\frac{\sqrt{2\pi}}{36864}\,c^{4}\nonumber\\
&&\hspace{50mm}
-\frac{2}{81081}\,c^{5}
-\frac{5\sqrt{2\pi}}{18579456}\,c^{6}
+\mathcal{O}(c^{7})\;.
\end{eqnarray}

The first two terms of (\ref{max[b-h] parabola small c}) are recovered immediately from the perturbative expansion $\max[b-h]\simeq\frac{\sqrt{2\pi}}{4}-\int_{0}^{1}\rmd x\,h(x)$ for general $h$. For the third term, using (\ref{mu2h[a]}) with the Fourier representation of the parabola $t(1-t)=\frac{1}{6}-\sum_{k\in\mathbb{Z}^{*}}\frac{\rme^{2\rmi\pi kt}}{2\pi^{2}k^{2}}$ gives the Schl\"omilch series \cite{W1995.1}
\begin{equation}
\label{Schlomilch}
\sum_{k=1}^{\infty}\frac{(-1)^{k}J_{1}(k\pi)}{k^{3}}=-\frac{\pi^{3}}{96}\;,
\end{equation}
which can be derived for instance by integrating (\ref{sum Bessel piecewise linear}) with respect to $a$, and leads to the term $\sqrt{2\pi}c^{2}/192$ in (\ref{max[b-h] parabola small c}). We observe that (\ref{Schlomilch}) can be recovered in a more direct (but non-rigorous) way by expanding the Bessel function as $J_{1}(k\pi)=\sum_{j=0}^{\infty}\frac{(-1)^{j}(k\pi/2)^{2j+1}}{j!(j+1)!}$, exchanging the sums over $j$ and $k$ and removing the divergent terms in the sum over $k$ with the prescription $\sum_{k=1}^{\infty}(-1)^{k}k^{2j-2}\equiv(2^{2j-1}-1)\,\zeta(2-2j)$. The identity (\ref{Schlomilch}) then follows from the explicit expressions $\zeta(2)=\pi^{2}/6$, $\zeta(0)=-1/2$ and $\zeta(2-2j)=0$ for $j\geq2$. Unfortunately, the same kind of reasoning does not allow to obtain the more general series (\ref{sum Bessel piecewise linear}).
\end{subsubsection}

\end{subsection}

\end{section}

\appendix
\begin{section}{Generating function of the height and eigenstates of TASEP}
\label{appendix sum eigenstates}
In this appendix, we recall how the generating function of the height function $H_{i}(\tm)$ of TASEP can be computed in terms of a deformed Markov operator.

We consider the vector space $V_{\Omega}$ with dimension $|\Omega|$ generated by the set of configurations $\Omega$. The elements $|\mathcal{C}\rangle$ of the canonical basis of $V_{\Omega}$ are noted either $|X_{1},\ldots,X_{N}\rangle$ or $|n_{1},\ldots,n_{L}\rangle$ in terms of the positions of the particles or the occupation numbers. The vector $|P(\tm)\rangle$, initially equal to $|P_{0}\rangle$, evolves in time by the master equation $\frac{\rmd}{\rmd\tm}|P(\tm)\rangle=M|P(\tm)\rangle$ with $M$ the Markov operator.

We consider the (globally) deformed Markov operator $M(\gamma)$ obtained after multiplying by $\rme^{\gamma/L}$ all non-diagonal elements of the Markov operator $M$ in the canonical basis. The operator $M(\gamma)$ commutes with the translation operator $U$ defined by $U|X_{1},\ldots,X_{N}\rangle=|X_{1}+1,\ldots,X_{N}+1\rangle$. It is then possible to diagonalize simultaneously $M(\gamma)$ and $U$. Introducing for $r=0,\ldots,|\Omega|-1$ the left and right eigenvectors $\langle\Psi_{r}(\gamma)|$ and $|\Psi_{r}(\gamma)\rangle$, one has
\begin{equation}
\langle\Psi_{r}(\gamma)|M(\gamma)=E_{r}(\gamma)\langle\Psi_{r}(\gamma)|
\qquad
M(\gamma)|\Psi_{r}(\gamma)\rangle=E_{r}(\gamma)|\Psi_{r}(\gamma)\rangle
\end{equation}
and
\begin{equation}
\label{Upsi}
\langle\Psi_{r}(\gamma)|U=\rme^{-\rmi p_{r}/L}\langle\Psi_{r}(\gamma)|
\qquad
U|\Psi_{r}(\gamma)\rangle=\rme^{-\rmi p_{r}/L}|\Psi_{r}(\gamma)\rangle\;.
\end{equation}
The dominant eigenstate $r=0$, corresponding for $\gamma\in\mathbb{R}$ to the eigenvalue of $M(\gamma)$ with largest real part, has momentum $p_{0}=0$. The eigenvalues $E_{r}(\gamma)$ are in general complex numbers, while the momenta $p_{r}$ are integer multiples of $2\pi$.

The master equation describing the evolution in time of the joint probability $P_{\tm}(\mathcal{C},Q_{i})$ of the microscopic state $\mathcal{C}$ and of the current $Q_{i}(\tm)$ couples the probabilities with different values of $Q_{i}$. This can be remedied by introducing the generating function $F_{\tm}(\mathcal{C},\gamma)=\sum_{Q_{i}\in\mathbb{Z}}\rme^{\gamma Q_{i}}P_{\tm}(\mathcal{C},Q_{i})$, with $\gamma$ a fugacity conjugate to $Q_{i}$. The master equation for $|F(\tm,\gamma)\rangle=\sum_{\mathcal{C}\in\Omega}F_{\tm}(\mathcal{C},\gamma)|\mathcal{C}\rangle$ is then $\frac{\rmd}{\rmd\tm}|F(\tm,\gamma)\rangle=M_{i}(\gamma)|F(\tm,\gamma)\rangle$ with $M_{i}(\gamma)$ a (local) deformation of the Markov operator obtained from $M$ after multiplying by $\rme^{\gamma}$ the elements of $M$ corresponding to moves from the site $i$ to the site $i+1$. The generating function of the current $\langle\rme^{\gamma Q_{i}(\tm)}\rangle$ can then be written as $\langle\rme^{\gamma Q_{i}(\tm)}\rangle=\sum_{\mathcal{C}\in\Omega}\langle\mathcal{C}|\rme^{\tm M_{i}(\gamma)}|P_{0}\rangle$. Globally and locally deformed Markov matrices have the same eigenvalues $E_{r}(\gamma)$ since they are related by the similarity transformation $M_{i}(\gamma)=\rme^{-\gamma S_{i}}M(\gamma)\rme^{\gamma S_{i}}$. The operator $S_{i}$, diagonal in the canonical basis, is such that $S_{i}|X_{1},\ldots,X_{N}\rangle=\frac{1}{L}\sum_{j=1}^{N}[X_{j}]_{i}$, where $[X]_{i}$ is the integer between $1$ and $L$ counting the position $X$ from site $i+1$, $[X+i]_{i}=X$ modulo $L$.

Since the height function is related to the current by $Q_{i}(\tm)=H_{i}(\tm)-H_{i}(0)$, the generating function of the height can be written as $\langle\rme^{\gamma H_{i}(\tm)}\rangle=\sum_{\mathcal{C}\in\Omega}\langle\mathcal{C}|\rme^{\tm M_{i}(\gamma)}\rme^{\gamma\mathsf{H0}_{i}}|P_{0}\rangle$, where the operator $\mathsf{H0}_{i}$ is defined by $\mathsf{H0}_{i}|n_{1},\ldots,n_{L}\rangle=\sum_{k=1}^{i}(\rhobar-n_{k})|n_{1},\ldots,n_{L}\rangle$. Using the identity $\mathsf{H0}_{i}=S_{0}-S_{i}$ and the similarity transform from $M_{i}(\gamma)$ to $M(\gamma)$, one has $\langle\rme^{\gamma H_{i}(\tm)}\rangle=\sum_{\mathcal{C}\in\Omega}\langle\mathcal{C}|\rme^{-\gamma S_{i}}\rme^{\tm M(\gamma)}\rme^{\gamma S_{0}}|P_{0}\rangle$. Expanding the operator $\rme^{\tm M(\gamma)}$ over the eigenstates of $M(\gamma)$ and using $S_{i}=U^{i}S_{0}U^{-i}$, (\ref{Upsi}) and $\sum_{\mathcal{C}\in\Omega}\langle\mathcal{C}|U^{i}=\sum_{\mathcal{C}\in\Omega}\langle\mathcal{C}|$ finally gives
\begin{equation}
\label{GFH[psi0]}
\langle\rme^{\gamma H_{i}(\tm)}\rangle=\sum_{r=0}^{|\Omega|-1}\frac{(\sum_{\mathcal{C}\in\Omega}\langle\mathcal{C}|\Psi_{r}^{0}(\gamma)\rangle)\,\langle\Psi_{r}^{0}(\gamma)|P_{0}\rangle}{\langle\Psi_{r}^{0}(\gamma)|\Psi_{r}^{0}(\gamma)\rangle}\;\rme^{\tm E_{r}(\gamma)+i\frac{\rmi p_{r}}{L}}\;,
\end{equation}
where $\langle\Psi_{r}^{0}(\gamma)|=\langle\Psi_{r}(\gamma)|\rme^{\gamma S_{0}}$ and $|\Psi_{r}^{0}(\gamma)\rangle=\rme^{-\gamma S_{0}}|\Psi_{r}(\gamma)\rangle$ are the left and right eigenvectors of the locally deformed Markov operator $M_{0}(\gamma)$ at site $0$ (modulo $L$). More generally, multiple point correlations can be considered by introducing the operator $M(\{\gamma_{i}\})$, obtained after multiplying by $\rme^{\gamma_{i}/L}$ the elements of $M$ corresponding to moves from the site $i$ to the site $i+1$ for any $i=1,\ldots,L$. The identity $M(\{\gamma_{i}\})=\rme^{-\frac{1}{L}\sum_{i=1}^{L}\gamma_{i}S_{i}}M(\overline{\gamma})\rme^{\frac{1}{L}\sum_{i=1}^{L}\gamma_{i}S_{i}}$ with $\overline{\gamma}=\frac{1}{L}\sum_{i=1}^{L}\gamma_{i}$ leads to
\begin{equation}
\fl\hspace{10mm}
\langle\rme^{\frac{1}{L}\sum_{i=1}^{L}\gamma_{i}H_{i}(\tm)}\rangle=\sum_{r=0}^{|\Omega|-1}\frac{(\sum_{\mathcal{C}\in\Omega}\langle\mathcal{C}|\rme^{\frac{1}{L}\sum_{i=1}^{L}\gamma_{i}\mathsf{H0}_{i}}|\Psi_{r}^{0}(\overline{\gamma})\rangle)\,\langle\Psi_{r}^{0}(\overline{\gamma})|P_{0}\rangle}{\langle\Psi_{r}^{0}(\overline{\gamma})|\Psi_{r}^{0}(\overline{\gamma})\rangle}\;\rme^{\tm E_{r}(\overline{\gamma})}\;.
\end{equation}

\end{section}

\begin{section}{Continuum limit of TASEP and Brownian bridges}
\label{appendix TASEP -> bb}
By Donsker's theorem, see e.g. \cite{D1999.1}, the height function $H_{i}$ becomes a standard Brownian bridge in the continuum when each configuration $\mathcal{C}\in\Omega$ of TASEP has the same weight. For completeness, we re-derive this property in \ref{section Donsker} by splitting the system into $M$ boxes of length $L/M$ with $1\ll M\ll L$. Then, in \ref{section bb master equation}, we compute the first correction in $L$, that is needed to derive (\ref{e[f bb]}) and (\ref{f bb}). Finally, the asymptotics of the operator $T(\gamma)$ defined in (\ref{T[H]}) is obtained in \ref{section T(gamma) asymptotics}.

\begin{subsection}{Sum over configurations of TASEP and Brownian bridges}
\label{section Donsker}
For a configuration $\mathcal{C}\in\Omega$ of TASEP with $L$ particles and $N$ sites, we partition the system into $M$ boxes of consecutive sites $B_{m}=\{(m-1)L/M+1,\ldots,mL/M\}$, $m=1,\ldots,M$. The integer $M$ has to divide $L$, and each box contains exactly $L/M$ sites.

We introduce the box variables
\begin{eqnarray}
\label{box rho_m[n_i]}
&& \rho_{m}=\frac{M}{L}\sum_{i\in B_{m}}n_{i}\\
\label{box sigma_m[rho_m]}
&& \sigma_{m}=\frac{\sqrt{L}(\rho_{m}-\rhobar)}{\sqrt{\rhobar(1-\rhobar)}}\\
\label{box h_m[sigma_m]}
&& h_{m}=-\frac{1}{M}\sum_{n=1}^{m}\sigma_{n}\;,
\end{eqnarray}
with $n_{i}$ the occupation numbers corresponding to $\mathcal{C}$. Since $n_{1}+\ldots+n_{L}=N$, one has $\rho_{1}+\ldots+\rho_{M}=\rhobar M$, $\sigma_{1}+\ldots+\sigma_{M}=0$ and $h_{0}=h_{M}=0$\footnote{The box variable $h_{0}$ in this section must not be confused with the height profile $h_{0}(x)$ from (\ref{h0[sigma0]}).}.

Let $f(h_{0},\ldots,h_{M})$ be an arbitrary function of the box variables $h_{m}$. Then, the average $\langle f(h_{0},\ldots,h_{M})\rangle_{\Omega}=|\Omega|^{-1}\sum_{\mathcal{C}\in\Omega}f(h_{0},\ldots,h_{M})$ over all configurations with the same weight is equal to
\begin{equation}
\label{box <f>[rho_m]}
\fl\hspace{20mm}
\langle f(h_{0},\ldots,h_{M})\rangle_{\Omega}=\frac{1}{|\Omega|}\,\sum_{\rho_{1},\ldots,\rho_{M}\in\frac{M}{L}[\![0,L/M]\!] \atop \rho_{1}+\ldots+\rho_{M}=\rhobar M}\!\!\!\!\!f(h_{0},\ldots,h_{M})\prod_{m=1}^{M}\C{L/M}{L\rho_{m}/M}\;,
\end{equation}
where the box variables $h_{m}$ on the right are defined from the box variables $\rho_{m}$ using (\ref{box h_m[sigma_m]}) and (\ref{box sigma_m[rho_m]}). The binomial coefficients represent the number of ways to place $L\rho_{m}/M$ particles among the $L/M$ sites of box $m$ with the exclusion constraint. Replacing the variables $\rho_{m}\in\frac{M}{L}[\![0,L/M]\!]$, $\rho_{1}+\ldots+\rho_{M}=\rhobar M$ with the variables $\sigma_{m}\in\frac{M}{\sqrt{\rhobar(1-\rhobar)L}}[\![-\frac{N}{M},\frac{L-N}{M}]\!]$, $\sigma_{1}+\ldots+\sigma_{M}=0$ defined in (\ref{box sigma_m[rho_m]}), the sum over the $\sigma_{m}$ is dominated at large $L$ by $\sigma_{m}$'s of order $L^{0}$. Using Stirling's formula (including the first few sub-leading terms), we have the large $L$ asymptotics with finite $\rhobar$ and $\sigma_{m}$
\begin{eqnarray}
&&
\fl\hspace{1mm}
\C{L/M}{L\rho_{m}/M}=\C{L/M}{\frac{L\rhobar}{M}(1+\frac{\sqrt{1-\rhobar}\,\sigma_{m}}{\sqrt{\rhobar}\,\sqrt{L}})}\\
&&
\fl\hspace{10mm}
\simeq
\frac{\rme^{-\frac{L}{M}(\rhobar\log\rhobar+(1-\rhobar)\log(\rhobar))-\frac{\sigma_{m}\sqrt{\rhobar(1-\rhobar)L}}{M}\,\log\frac{\rhobar}{1-\rhobar}-\frac{\sigma_{m}^{2}}{2M}}}{\sqrt{2\pi\rhobar(1-\rhobar)L/M}}
\Big(1-\frac{(1-2\rhobar)\sigma_{m}(3-\frac{\sigma_{m}^{2}}{M})}{6\sqrt{\rhobar(1-\rhobar)L}}
+\frac{B(\sigma_{m})}{L}\Big)\nonumber
\end{eqnarray}
and
\begin{equation}
\C{L}{\rhobar L}
\simeq
\frac{\rme^{-L(\rhobar\log\rhobar+(1-\rhobar)\log(\rhobar))}}{\sqrt{2\pi\rhobar(1-\rhobar)L}}\,\Big(1-\frac{B}{L}\Big)\;.
\end{equation}
The precise values of the coefficients $B(\sigma_{m})$ and $B$ will not be needed in the following. Using $\sigma_{1}+\ldots+\sigma_{M}=0$, one finds after some simplifications
\begin{eqnarray}
&&
\fl\hspace{10mm}
\frac{\prod\limits_{m=1}^{M}\displaystyle\C{L/M}{\frac{L\rhobar}{M}(1+\frac{\sqrt{1-\rhobar}\,\sigma_{m}}{\sqrt{\rhobar}\,\sqrt{L}})}}{\displaystyle\C{L}{\rhobar L}}
\simeq\frac{M^{M/2}\rme^{-\frac{1}{2M}\sum\limits_{m=1}^{M}\sigma_{m}^{2}}}{(2\pi\rhobar(1-\rhobar)L)^{\frac{M-1}{2}}}\,\Big(1+\frac{(1-2\rhobar)\sum_{m=1}^{M}\sigma_{m}^{3}}{6M\sqrt{\rhobar(1-\rhobar)L}}\\
&&\fl\hspace{105mm}
+\frac{B(\sigma_{1},\ldots,\sigma_{M})}{L}\Big)\;.\nonumber
\end{eqnarray}

The large $L$ asymptotics of (\ref{box <f>[rho_m]}) can be computed using the Euler-Maclaurin formula. Since the summand in (\ref{box <f>[rho_m]}) is exponentially smaller for $\sigma_{m}$ close to the boundaries of the summation range than for $\sigma_{m}\sim L^{0}$, boundary terms of the Euler-Maclaurin formula with Bernoulli numbers do not contribute to the algebraic part in $L$ of the expansion. We transform only the sums over $\sigma_{1}$, \ldots, $\sigma_{M-1}$ into integrals, since $\sigma_{M}$ is then fixed by the constraint $\sigma_{1}+\ldots+\sigma_{M}$. In the end, we add another integral for $\sigma_{M}$ and a $\delta$ function in order to enforce the constraint. One has
\begin{eqnarray}
&&\fl
\langle f(h_{0},\ldots,h_{M})\rangle_{\Omega}\simeq\frac{M\sqrt{2\pi}}{(2\pi M)^{M/2}}\int_{-\infty}^{\infty}\rmd\sigma_{1}\ldots\rmd\sigma_{M}\delta(\sigma_{1}+\ldots+\sigma_{M})f(h_{0},\ldots,h_{M})\\
&&
\fl\hspace{45mm}
\times\rme^{-\frac{1}{2M}\sum_{m=1}^{M}\sigma_{m}^{2}}\Big(1+\frac{(1-2\rhobar)\sum_{m=1}^{M}\sigma_{m}^{3}}{6M\sqrt{\rhobar(1-\rhobar)L}}+\frac{B(\sigma_{1},\ldots,\sigma_{M})}{L}\Big)\;,\nonumber
\end{eqnarray}
where the $h_{m}$ are given in terms of the $\sigma_{m}$ by (\ref{box h_m[sigma_m]}). After a change of variables from the $\sigma_{m}$ to the $h_{m}$, we finally obtain
\begin{eqnarray}
&&\fl\hspace{10mm}
\langle f(h_{0},\ldots,h_{M})\rangle_{\Omega}\simeq\frac{\sqrt{2\pi}}{(2\pi/M)^{M/2}}\int_{-\infty}^{\infty}\rmd h_{0}\ldots\rmd h_{M}\delta(h_{0})\delta(h_{M})f(h_{0},\ldots,h_{M})\\
&&\fl\hspace{10mm}
\times\rme^{-\frac{M}{2}\sum_{m=1}^{M}(h_{m}-h_{m-1})^{2}}\Big(1-\frac{(1-2\rhobar)M^{2}\sum_{m=1}^{M}(h_{m}-h_{m-1})^{3}}{6\sqrt{\rhobar(1-\rhobar)L}}+\frac{C(h_{0},\ldots,h_{M})}{L}\Big)\;,\nonumber
\end{eqnarray}
where the precise expression for $C(h_{0},\ldots,h_{M})$ will not be needed in the following. We recognize in the expression above the $M$-point distribution of the standard Brownian bridge $b$:
\begin{eqnarray}
&&\fl\hspace{5mm}
\langle f(b(0),b(1/M),\ldots,b(1))\rangle_{b}\\
&&\fl\hspace{15mm}
=\frac{\sqrt{2\pi}}{(2\pi/M)^{M/2}}\int_{-\infty}^{\infty}\rmd h_{0}\ldots\rmd h_{M}\delta(h_{0})\delta(h_{M})f(h_{0},\ldots,h_{M})\rme^{-\frac{M}{2}\sum_{m=1}^{M}(h_{m}-h_{m-1})^{2}}\;,\nonumber
\end{eqnarray}
which can be derived for instance from the representation $b(x)=w(x)-xw(1)$ where $w$ is the Wiener process starting at $w(0)=0$. Thus, one has
\begin{equation}
\label{box <f>[b] leading order}
\langle f(h_{0},\ldots,h_{M})\rangle_{\Omega}\underset{L\to\infty}{\to}\langle f(b(0),b(1/M),\ldots,b(1))\rangle_{b}\;.
\end{equation}
This is essentially equivalent to Donsker's theorem, and sufficient to establish (\ref{theta bb}). The large $L$ expansion up to order $1$ in $L$ is however needed in order to derive (\ref{e[f bb]}), (\ref{f bb}). One has
\begin{eqnarray}
\label{box <f>[b] correction}
&&\fl\hspace{5mm} \langle f(h_{0},\ldots,h_{M})\rangle_{\Omega}\simeq\Big\langle f(b(0),b(1/M),\ldots,b(1))\\
&&\fl\hspace{15mm} \times\Big(1+\frac{-\frac{1-2\rhobar}{6}\,M^{2}\sum_{m=1}^{M}(b(\frac{m}{M})-b(\frac{m-1}{M}))^{3}}{\sqrt{\rhobar(1-\rhobar)L}}+\frac{C(b(0),b(1/M),\ldots,b(1))}{L}\Big)\Big\rangle_{b}\nonumber\;.
\end{eqnarray}
The expansion up to first order in $L$ of $\langle\openone_{\{n_{1}=0\}}\openone_{\{n_{L}=1\}}f(h_{0},\ldots,h_{M})\rangle_{\Omega}$ is also needed in order to treat (\ref{eigenvalue eq[T]}). The calculation is essentially the same except for a small modification in the binomial coefficients counting the number of ways to place the particles in the first and the last box:
\begin{eqnarray}
&&\fl
\langle\openone_{\{n_{1}=0\}}\openone_{\{n_{L}=1\}}f(h_{0},\ldots,h_{M})\rangle_{\Omega}=\frac{1}{|\Omega|}\,\sum_{\rho_{1},\ldots,\rho_{M}\in\frac{M}{L}[\![0,L/M]\!] \atop \rho_{1}+\ldots+\rho_{M}=\rhobar M}\!\!\!\!\!f(h_{0},\ldots,h_{M})\\
&& \times\C{L/M-1}{L\rho_{1}/M}\C{L/M-1}{L\rho_{M}/M-1}\prod_{m=2}^{M-1}\C{L/M}{L\rho_{m}/M}\;.
\end{eqnarray}
Using
\begin{equation}
\fl\hspace{5mm}
\C{L/M-1}{L\rho_{1}/M}=(1-\rho_{1})\C{L/M}{L\rho_{1}/M}
\quad\text{and}\quad
\C{L/M-1}{L\rho_{M}/M-1}=\rho_{M}\C{L/M}{L\rho_{M}/M}\;,
\end{equation}
similar calculations as above lead to
\begin{eqnarray}
&&\fl\hspace{5mm}
\langle\openone_{\{n_{1}=0\}}\openone_{\{n_{L}=1\}}f(h_{0},\ldots,h_{M})\rangle_{\Omega}\simeq\rhobar(1-\rhobar)\Big\langle f(b(0),b(1/M),\ldots,b(1))\\
&&\fl\hspace{10mm}
\times\Big(1+\frac{-\frac{1-2\rhobar}{6}\,M^{2}\sum_{m=1}^{M}(b(\frac{m}{M})-b(\frac{m-1}{M}))^{3}+(1-\rhobar)b(1-\frac{1}{M})+\rhobar Mb(\frac{1}{M})}{\sqrt{\rhobar(1-\rhobar)L}}\nonumber\\
&&\fl\hspace{25mm}
+\frac{C(b(0),b(1/M),\ldots,b(1))+M^{2}b(\frac{1}{M})b(1-\frac{1}{M})}{L}\nonumber\\
&&\fl\hspace{30mm}
+\frac{-\frac{1-2\rhobar}{6}\,M^{3}((1-\rhobar)b(1-\frac{1}{M})+\rhobar b(\frac{1}{M}))\sum_{m=1}^{M}(b(\frac{m}{M})-b(\frac{m-1}{M}))^{3}}{L}\Big)\Big\rangle_{b}\nonumber\;,
\end{eqnarray}
where $C(b(0),b(1/M),\ldots,b(1))$ is the same as in (\ref{box <f>[b] correction}).
\end{subsection}

\begin{subsection}{Dominant eigenvalue at large $L$ in terms of Brownian bridges}
\label{section bb master equation}
We want to take the large $L$ limit of (\ref{eigenvalue eq[T]}) for fixed $\rhobar$ and $s$ with $\gamma=s/\sqrt{\rhobar(1-\rhobar)L}$. Inserting the decomposition of the identity between the operators $T(\gamma)$ and using (\ref{box <f>[b] correction}), (\ref{T[b]}) and
\begin{equation}
\frac{E_{0}(\gamma)}{\rme^{\gamma}-1}\simeq\rhobar(1-\rhobar)\Big(1-\frac{s}{2\sqrt{\rhobar(1-\rhobar)L}}+\frac{\frac{s^{2}}{12\rhobar(1-\rhobar)}+\frac{e(s)}{s}}{L}\Big)\;,
\end{equation}
we obtain at half filling $\rhobar=1/2$ the relations (\ref{e[f bb]}) and
\begin{equation}
\fl\hspace{5mm} \frac{\langle(b_{n}'(0)-b_{n}'(1))\,\rme^{-s\max[b_{1}-h_{0}]-s\sum_{j=2}^{n}\max[b_{j}-b_{j-1}]}\rangle_{b_{1},\ldots,b_{n}}}{\langle\rme^{-s\max[b_{1}-h_{0}]-s\sum_{j=2}^{n}\max[b_{j}-b_{j-1}]}\rangle_{b_{1},\ldots,b_{n}}}=-s+\mathcal{O}(s^{n})
\end{equation}
The derivatives $b_{n}'(0)$ and $b_{n}'(1)$ are understood as the limits $b_{n}'(0)=\lim_{M\to\infty}Mb_{n}(1/M)$, $b_{n}'(1)=-\lim_{M\to\infty}Mb_{n}(1-1/M)$. At arbitrary $\rhobar$, we obtain additionally
\begin{equation}
\fl\hspace{2mm} \frac{\langle(b_{n}'(0)+b_{n}'(1))(\int_{0}^{1}\rmd x\,b_{n}'(x)^{3})\,\rme^{-s\max[b_{1}-h_{0}]-s\sum_{j=2}^{n}\max[b_{j}-b_{j-1}]}\rangle_{b_{1},\ldots,b_{n}}}{\langle\rme^{-s\max[b_{1}-h_{0}]-s\sum_{j=2}^{n}\max[b_{j}-b_{j-1}]}\rangle_{b_{1},\ldots,b_{n}}}=s^{2}+\mathcal{O}(s^{n})\;,
\end{equation}
where the integral is understood as $\int_{0}^{1}\rmd x\,b_{n}'(x)^{3}\underset{M\to\infty}{=}M^{2}\sum_{m=1}^{M}(b_{n}(\frac{m}{M})-b_{n}(\frac{m-1}{M}))^{3}$.
\end{subsection}

\begin{subsection}{$T(\gamma)$ between typical configurations}
\label{section T(gamma) asymptotics}
We consider two typical configurations $\mathcal{C}_{1}$ and $\mathcal{C}_{2}$ with respective associated density profiles $\rhobar+\sqrt{\rhobar(1-\rhobar)}\,\sigma_{1}(x)/\sqrt{L}$ and $\rhobar+\sqrt{\rhobar(1-\rhobar)}\,\sigma_{2}(x)/\sqrt{L}$. The fugacity $\gamma$ is taken equal to $\gamma=s/\sqrt{\rhobar(1-\rhobar)L}$, and we are interested in the large $L$ limit of $\langle\mathcal{C}_{2}|T(\gamma)|\mathcal{C}_{1}\rangle$ with fixed $\rhobar$ and $s$.

We partition again the system in $M$ boxes of size $L/M$ and introduce the box variables $h_{m,1}$ and $h_{m,2}$ corresponding respectively to $\mathcal{C}_{1}$ and $\mathcal{C}_{2}$ as in (\ref{box h_m[sigma_m]}). From (\ref{T[H]}), one has at large $M$ (with $M\ll L$)
\begin{equation}
\langle\mathcal{C}_{2}|T(\gamma)|\mathcal{C}_{1}\rangle\simeq\rme^{-s\max\limits_{0\leq m\leq M}[h_{m,2}-h_{m,1}]}\;.
\end{equation}
Defining now the height profiles $h_{1}(x)$ and $h_{2}(x)$ from $\sigma_{1}$ and $\sigma_{2}$ as in (\ref{h0[sigma0]}), we finally obtain
\begin{equation}
\label{T[b]}
\langle\mathcal{C}_{2}|T(\gamma)|\mathcal{C}_{1}\rangle\underset{1\ll M\ll L}{\simeq}\rme^{-s\max\limits_{0\leq x\leq 1}[h_{2}(x)-h_{1}(x)]}
\end{equation}
\end{subsection}

\end{section}

\end{document}